\documentclass[aps,prfluids,superscript,address,
longbibliography,
unsortedaddress,
amsmath,
amssymb]{revtex4-2}

\usepackage{float}
\usepackage{graphicx}
\usepackage{subcaption}
\usepackage{siunitx}
\usepackage{empheq}
\usepackage[french]{babel}
\usepackage[T1]{fontenc}
\usepackage{mathtools}

\newsavebox{\largestimage}

\newcommand{\iemn}{Univ. Lille, CNRS, Centrale Lille, Univ. Polytechnique Hauts-de-France, UMR 8520 - IEMN - Institut d'Electronique de Microélectronique et de Nanotechnologie, F-59000 Lille, France}

\newcommand{\We}{\mathrm{We}}
\newcommand{\Weloc}{\We_\mathrm{loc}}
\renewcommand{\Re}{\mathrm{Re}}
\newcommand{\degC}{\degreeCelsius}
\newcommand{\Frad}{F_\mathrm{rad}}
\newcommand{\Fup}{F_\mathrm{up}}
\newcommand{\dl}{\delta_\ell}

\usepackage{ragged2e}

\usepackage{xcolor}
\usepackage{ulem}

\makeatletter
\renewcommand{\p@section}{\thesection\expandafter\@gobble}
\renewcommand{\p@subsection}{\thesection\expandafter\@gobble}
\makeatother

\makeatletter
\renewcommand\@make@capt@title[2]{%
 \@ifx@empty\float@link{\@firstofone}{\expandafter\href\expandafter{\float@link}}%
  {\textbf{#1}}\@caption@fignum@sep#2\quad}%
\makeatother

\begin{document}


\title{Vertical impact of a water jet on a hot plate: from a growing drop to spray formation.}


\author{A. Goerlinger}
 \email{aurelien.goerlinger@univ-lille.fr}
\author{A. Germa}
\author{F. Zoueshtiagh}
\author{A.Duchesne}
\affiliation{\iemn}


\date{\today}

\begin{abstract}
In this article, we experimentally investigate the impact of a submillimetric water jet on a horizontal surface heated well above the "static" Leidenfrost temperature of water. We observe the transition from a regime where a single drop grows at the impingement point to a regime of spray formation. The main control parameter appears to be the jet Weber number ($\We$). The first regime persists until $\We \lesssim 30$ whereas the spray formation occurs for $\We \gtrsim 40$. Surprisingly, we found no influence of the hot plate's temperature on the reported phenomena.
We particularly focus on the second regime, where the liquid jet spreads on the plate, forming a liquid sheet that eventually lifts off and breaks into droplets. We characterized this regime by the radius $r_c$ of the liquid sheet when it is still in contact with the plate and the angle of ejection $\theta$ of the droplets. We further examine the ejected droplets by characterizing their speed and sizes.
Simple models are proposed to predict the dependencies and order of magnitudes of $r_c$ and $\theta$. We also aim to predict the critical Weber number at which the transition between the two regimes occurs. Our models show reasonable agreement with our experimental data. Finally, we compare the energy transferred from the jet to the droplets with results reported in the literature for impacts on unheated surfaces, finding a difference of nearly a factor 2.
\end{abstract}


\maketitle

\section{Introduction}

When a drop of liquid is placed on a surface heated well above the boiling point of the liquid, the drop's lifetime becomes significantly greater than what can be expected. This phenomenon is due to the formation of a vapor layer between the droplet and the hot plate, hindering efficient heat transfer. This phenomenon was initially observed by Leidenfrost in 1756 and has been named after him ever since \cite{leidenfrostFixationWaterDiverse1966}. Because of its importance in many industrial applications such as cooling systems \cite{raudenskySecondaryCoolingContinuous2005,ngSuppressionLeidenfrostEffect2015} or sprays \cite{isakovExploringLeidenfrostEffect2017}, the Leidenfrost effect has been the topic of extensive research in the case of a sessile drop on a hot plate \cite{bernardinLeidenfrostPointExperimental1999}.


Other research delved into the dynamics of a drop with an initial vertical velocity colliding with a hot plate \cite{quereLeidenfrostDynamics2013,liangReviewDropImpact2017,caiReviewDynamicLeidenfrost2023}. For instance, it was observed that the drop can spread upon impact~\cite{lastakowskiBridgingLocalGlobal2014,ribouxMaximumDropRadius2016}, elastically bounce off the heated surface,  \cite{bianceElasticityInertialLiquid2006},  or generate a spray~\cite{bianceDropFragmentationDue2011,ribouxMaximumDropRadius2016,vanlimbeekOriginSprayFormation2017}. The existence of different boiling regimes as well as impact patterns have also been investigated \cite{bianceDropFragmentationDue2011,tranDropImpactSuperheated2012,khavariFingeringPatternsDroplet2015}.


However, unlike the impact of a drop, little is known about jet impingement on a heated surface beyond the classical industrial applications of quenching. Studies in this area are primarily engineering-focused, centering on the heat transfer between the liquid jet and the plate, and distinguish two main areas: a wet area near the jet's impact point, where the liquid is in contact with the hot plate, and a dry area far from the impact point \cite{agrawalSurfaceQuenchingJet2019}. These investigations cover planar~\cite{vaderConvectiveBoilingHeat1988,wolfTurbulentDevelopmentFree1993,robidouControlledCoolingHot2002,robidouLocalHeatTransfer2003,bogdanicTwophaseStructureHot2009} or circular~\cite{koberleTemperatureControlledMeasurements1994,baongaExperimentalStudyHydrodynamic2006,leocadioHeatTransferBehavior2009,karwaHydrodynamicModelSubcooled2011} jets, considering horizontal as well as inclined surfaces \cite{maLocalCharacteristicsImpingement1997a}. 
They mostly propose empirical modelings of the heat transfer in a transient regime where the temperature of the surface is not held constant, and very few of these works include a hydrodynamic model~\cite{baongaExperimentalStudyHydrodynamic2006,karwaHydrodynamicModelSubcooled2011}.

To remedy this lack of hydrodynamic descriptions, we explore in this study the scenario of a circular jet impacting a hot surface with a temperature significantly exceeding the boiling point of the liquid. We will demonstrate that upon impact, the jet can atomize, and under the tested experimental conditions of a sub-millimetric jet and a constant temperature maintained on the heated surface, the system exhibits quasi-steady behavior. Special attention will be given to the hydrodynamic characterization of the liquid as the jet impacts the hot plate and the conditions that lead to droplet ejections. We explore the system behaviour for varying surface temperatures, jet radii, and different flow rates. We employ simplified models to derive scaling laws consistent with experimental observations and characterize the transition that leads to this regime. Finally, we examine the velocity and radius of the ejected droplets.

\section{Experimental apparatus}
\begin{figure}[H]
    \centering
    \includegraphics[width=.5\linewidth]{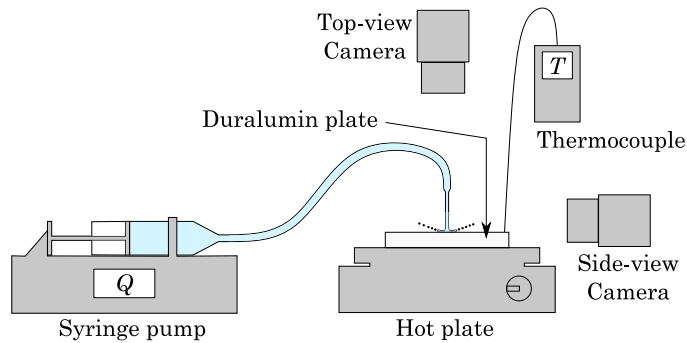}
    \caption{Experimental apparatus. The thermocouple indicates the temperature $T$ of the surface of the Duralumin disk. The syringe pump enables us to control the flowrate $Q$ of the jet.}
    \label{fig:setup}
\end{figure}

Our experimental apparatus is shown Fig.~\ref{fig:setup}. All experiments were conducted with deionized water. A syringe pump (Harvard Apparatus Pump 11 Elite) mounted with a 50 mL glass syringe provided flows of desired flowrates, $Q$, through a needle of radius~$a$ ranging from 85~\si{\micro\metre} to 250\si{\micro\metre}. Flow rates ranging from 0.075 mL/s to 1.55 mL/s were employed depending on the needle diameter. The Reynolds number in the injector, defined as $\mathrm{Re}=Q/\nu a$ with $\nu$ the kinematic viscosity of water, is estimated to range between 900 and 6700.
The water jet impacts a Duralumin disk with a thickness of 1.1 cm and a diameter of 16 cm, positioned 2 cm below the needle. The Duralumin disk is heated using a Fisherbrand Isotemp hot plate, enabling us to raise the plate's temperature from 300\si{\degC} up to 500\si{\degC} with a 600 W heating power. The surface temperature $T$ is monitored via a K-type thermocouple with a 0.1 K accuracy. Experiments feature a side-view via a Mikrotron Motionblitz Cube 4 camera mounted with a Tamron SP AF28-75mm camera lens, and a top-view via a Phantom Miro C211 camera mounted with a Sigma 105~mm 1:2.8 DG Macro camera lens. 

\vspace{0.1cm}

\noindent Side-view acquisitions are made at 150 FPS whereas top-view acquisitions are made at much higher FPS, between 3300 FPS and 6800 FPS. Lighting is provided by 110 W LED flood lights. Data were analyzed via Python, Matlab and ImageJ softwares.

We performed initial experiments to verify that the outcomes were unaffected by the precise location of the impinging jet, thereby ensuring uniform temperature distribution across the heated disk. Following each experiment, we noted a small temporary decrease in the temperature of the disk caused by the impact of the water jet. To ensure the plate's temperature returned to its initial value before subsequent jet impacts, various waiting times between each run were tested, yielding consistent results. Finally, our observations remain consistent throughout the entire experiment duration. Moreover, the recorded temperature change on the disk never surpasses 2\% of the disk temperature setting.  This indicates that our experiments are in a steady-state condition.

\section{Results}

We explore the behavior of the system by analyzing the impact of temperature in relation to the Weber number, $\We$. This parameter, which compares the inertia to surface tension effects, is defined as:
\begin{equation}
    \We=\frac{2\rho u_\mathrm{jet}^2 a}{\gamma}=\frac{2\rho Q^2}{\pi^2 \gamma a^3}
    \label{eq:weber}
\end{equation}
where $u_\mathrm{jet}$ is the jet's flow velocity, $a$ is the radius of the jet, $\rho=1000$ \si{\kilo\gram\per\cubic\metre} is the density of water at 20\si{\degC} and $\gamma=73$ \si{\milli\newton\per\metre} is the surface tension of water at 20\si{\degC}. In the present experiments, $\We$ ranges from 30 to 660 with the impinging jet remaining continuous.

\begin{figure}[H]
\hspace{1cm}
    \begin{subfigure}[t]{.45\linewidth}
        \centering
        \includegraphics[width=0.55\textwidth]{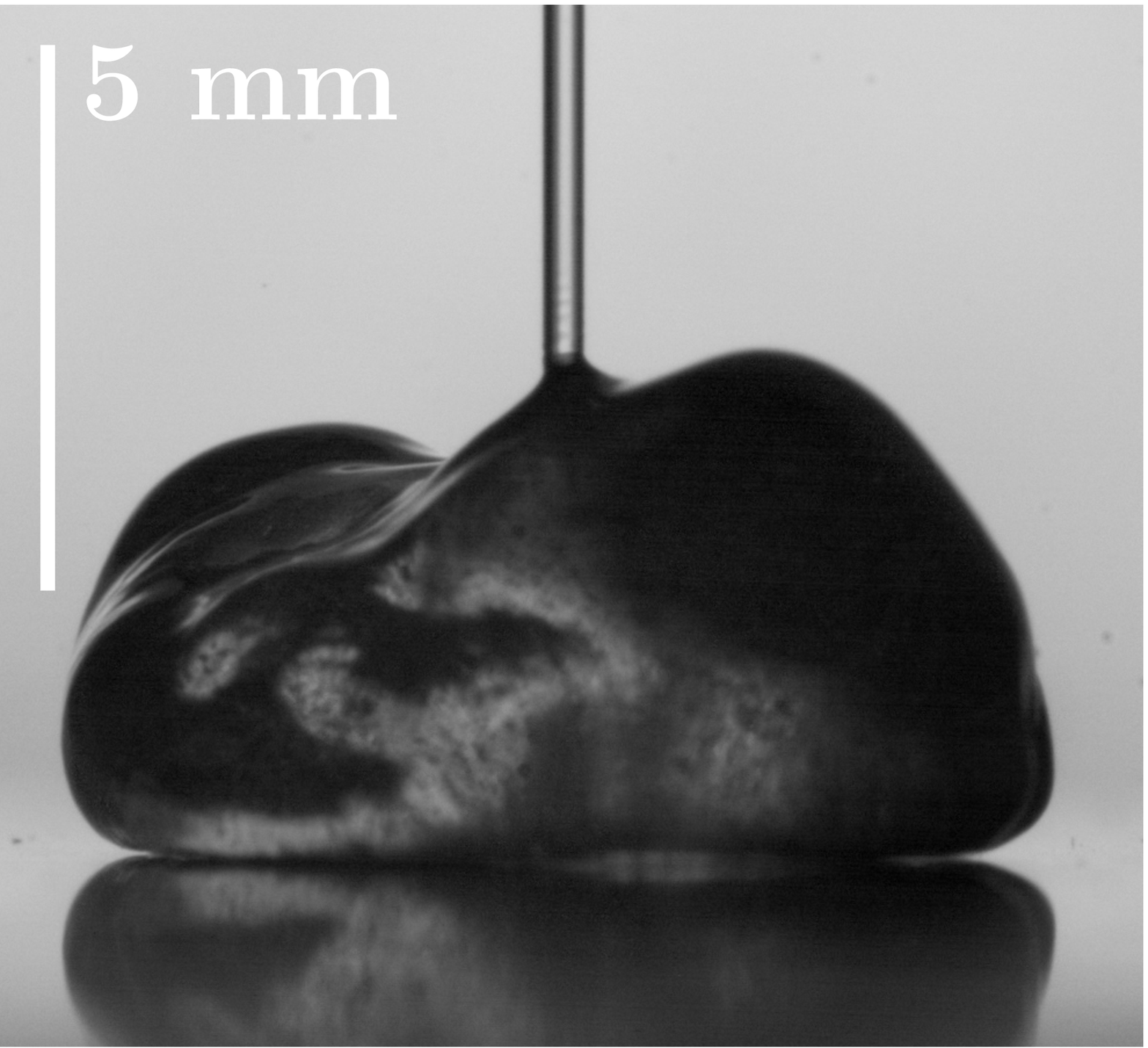}
        \subcaption{$\We \simeq 24$} 
        \label{subfig:gouttes}
    \end{subfigure}\hfill
    \begin{subfigure}[t]{.45\linewidth}
        \centering
        \includegraphics[width=0.55\textwidth]{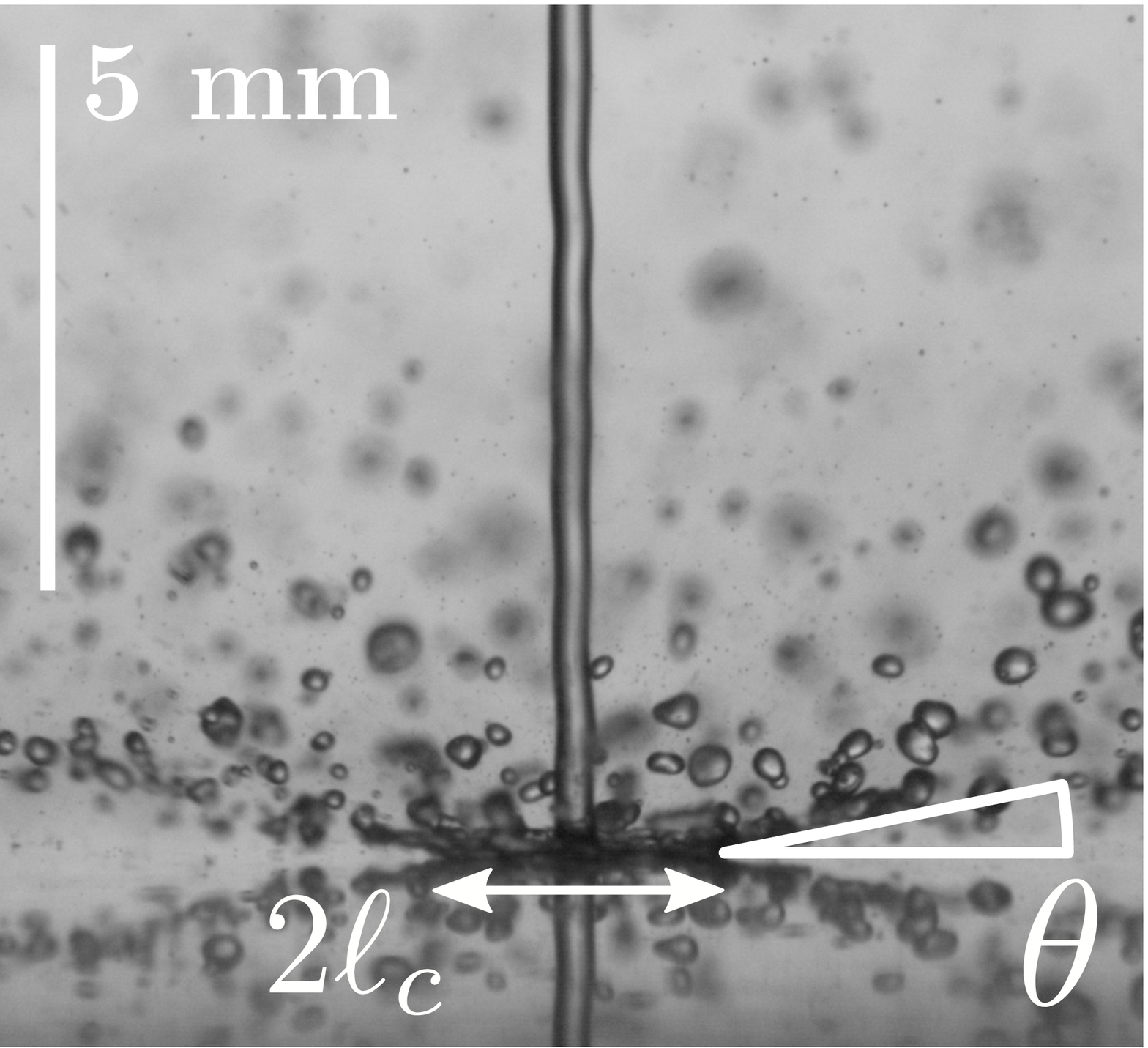}
        \subcaption{ $\We \simeq 98$} 
        \label{subfig:gouttelettes}
    \end{subfigure}
    \caption{\justifying Impact of a jet of water on a hot Duralumin disk for two different Weber numbers of the jet. The radius of the jet is $a=195$ \si{\micro\metre}, the surface temperature of the Duralumin disk is $T=350$\si{\degC}. (a) Growing drop regime. A single drop remains beneath the jet, enlarges, and eventually separates from the jet, making room for a new drop to form. (b) Droplet ejection regime. A circular liquid pad with a radius $r_c$ forms beneath the jet and fractures at its periphery into small droplets that are expelled at an angle $\theta$ relative to the horizontal.}
    \label{fig:regimes}
\hspace{1cm}
\end{figure}

Two primary regimes, namely (i) and (ii) were identified in the experiments. Fig.~\ref{fig:regimes} shows the characteristic shapes of the jet upon its impact with the hot plate for regimes (i) and (ii), respectively. Regime (i) is associated with $\We \lesssim 30$ where a single drop stays under the jet and grows until detaches from the jet and a new drop starts forming (see Fig.~\ref{subfig:gouttes} and Supplemental Material \cite{SuppMat} M1). Here, the drop's behavior under the jet recalls the classical Leidenfrost effect. When $\We$ exceeds approximately 40, the regime (ii) occurs, \textit{i.e.} the jet striking the hot plate results in the formation of a roughly circular liquid sheet that rapidly breaks up into small droplets, which are ejected at a characteristic angle $\theta$ (see Fig.~\ref{subfig:gouttelettes} and Supplemental Material \cite{SuppMat} M2). In this regime the downward pressure from the impinging jet results in direct contact between
the water and the heated surface of the disk, thus inhibiting any Leidenfrost effect near the point of jet impact. This was corroborated by the presence of water vapor emission, accompanied by the distinctive sizzling noise associated with boiling water. Furthermore, Karwa \textit{et al.}'s formulation~\cite{karwaHydrodynamicModelSubcooled2011} for the minimum temperature at which film boiling can occur beneath a jet indicates 840\si{\degreeCelsius}, a value significantly higher than the temperatures investigated in this study. For intermediary Weber numbers ($30 \lesssim \We \lesssim 40$), we observe a transition state in which features of both regimes can be observed. The emergence of these regimes in relation to the Weber number (\(\We\)) remains mostly unaffected by changes in plate temperature or jet radius.

In regime (ii) two characteristics, the ejection angle $\theta$ of the droplets with respect to the horizontal and the radius $r_c$ of the liquid sheet while it is still in contact with the hot plate, were investigated. Figure~\ref{fig:photos_differents_weber} displays images of experiments conducted at a fixed jet radius but with varying Weber numbers. It can be observed that the ejection angle $\theta$ decreases as the Weber number increases. This seems somewhat counterintuitive, as one might naturally expect that a stronger jet would lead to a higher angle, as is typically observed when a drop impacts an unheated surface~\cite{burzynskiSplashingHighspeedDrops2020}. 
Furthermore, the radius $r_c$ is observed to increase with the Weber number.
\noindent\begin{minipage}[t]{0.48\linewidth}
    \begin{figure}[H]
        \centering
        \includegraphics[width=\linewidth]{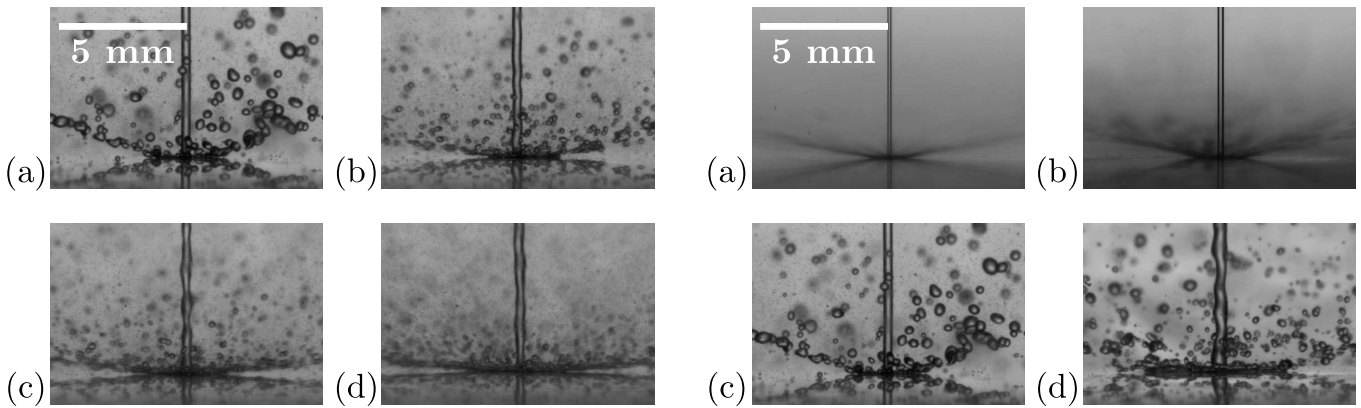}
        \caption{\justifying Droplet ejection regime observed when a jet of radius $a=195$ \si{\micro\metre} impact a Duralumin disk heated at $T=350$\si{\degC} for different Weber numbers $\We$. (a) $\We=47$. (b) $\We=100$. (c) $\We=190$. (d) $\We=560$. As $\We$ increases, one can observe the ejection angle $\theta$ flattening and the radius $r_c$ of the contact area widening. One can also note that the jet starts to become unstable for Weber numbers above 200.}
        \label{fig:photos_differents_weber}
    \end{figure}
\end{minipage}\hfill
\begin{minipage}[t]{0.48\linewidth}
    \begin{figure}[H]
        \centering
        \includegraphics[width=\linewidth]{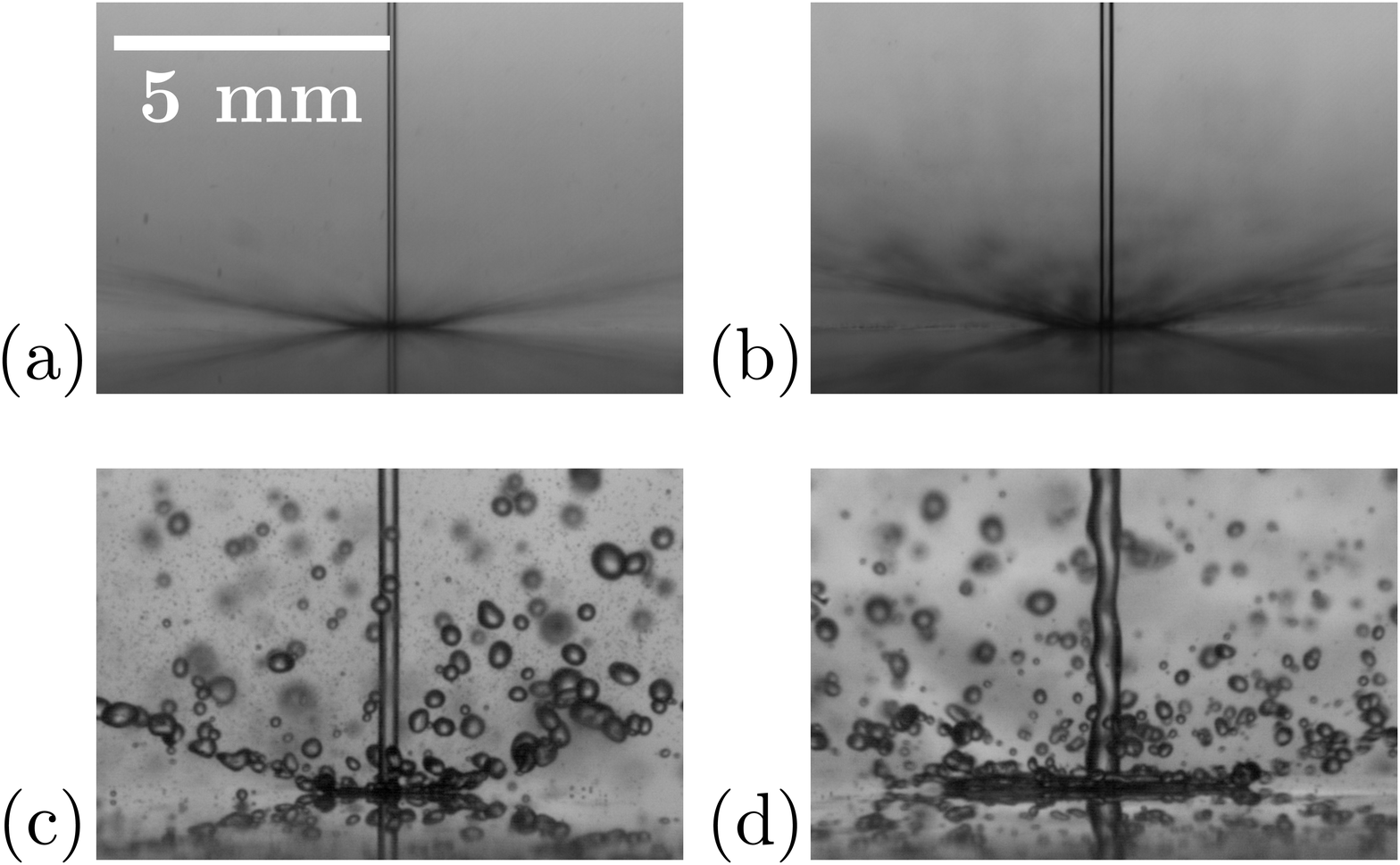}
        \caption{\justifying Droplet ejection regime observed when a jet of Weber number $\We\simeq 185$ impacts a Duralumin disk heated at $T=350$\si{\degC} for different jet radii $a$. (a) $a=85$ \si{\micro\metre}. (b) $a=125$ \si{\micro\metre}. (c) $a=195$ \si{\micro\metre}. (d) $a=250$ \si{\micro\metre}. As $a$ increases, one can observe the radius $r_c$ of the contact area widening, whereas the ejection angle $\theta$ does not seem to depend on the jet radius $a$.}
        \label{fig:photos_differents_aiguilles}
    \end{figure}
\end{minipage}

\noindent Additionally, $r_c$ also increases with the jet radius $a$, as demonstrated in Fig.~\ref{fig:photos_differents_aiguilles}, unlike the ejection angle $\theta$, which remains constant with respect to $a$. It is noteworthy that for Weber numbers exceeding approximately 200, instabilities begin to emerge in the jet. We attribute these instabilities to corrugation generated by the turbulent nature of the jet at high Reynolds number ($\Re \gtrsim 3000$).

\begin{figure}[H]
    \centering
    \includegraphics[width=.5\linewidth]{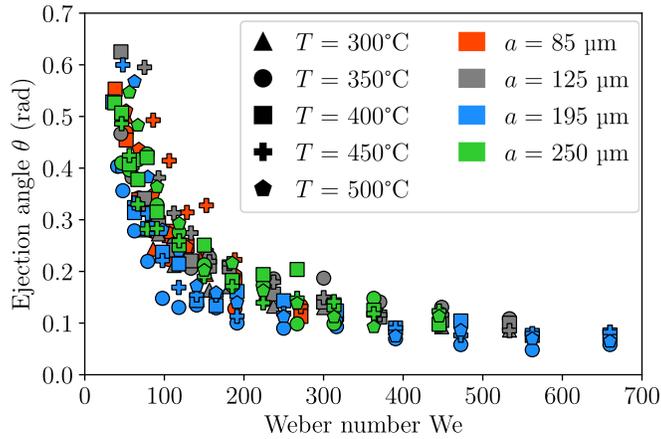}
    \caption{\justifying Ejection angle $\theta$ plotted against the Weber number $\We$ for various surface temperatures $T$ and jet radii $a$. The data illustrates that $\theta$ solely varies with $\We$, without being influenced by either $T$ or $a$.}
    \label{fig:theta}
\end{figure}

\begin{figure}[H]
    \centering
    \begin{subfigure}[t]{.5\linewidth}
        \centering
        \vskip 0pt
        \includegraphics[width=\textwidth]{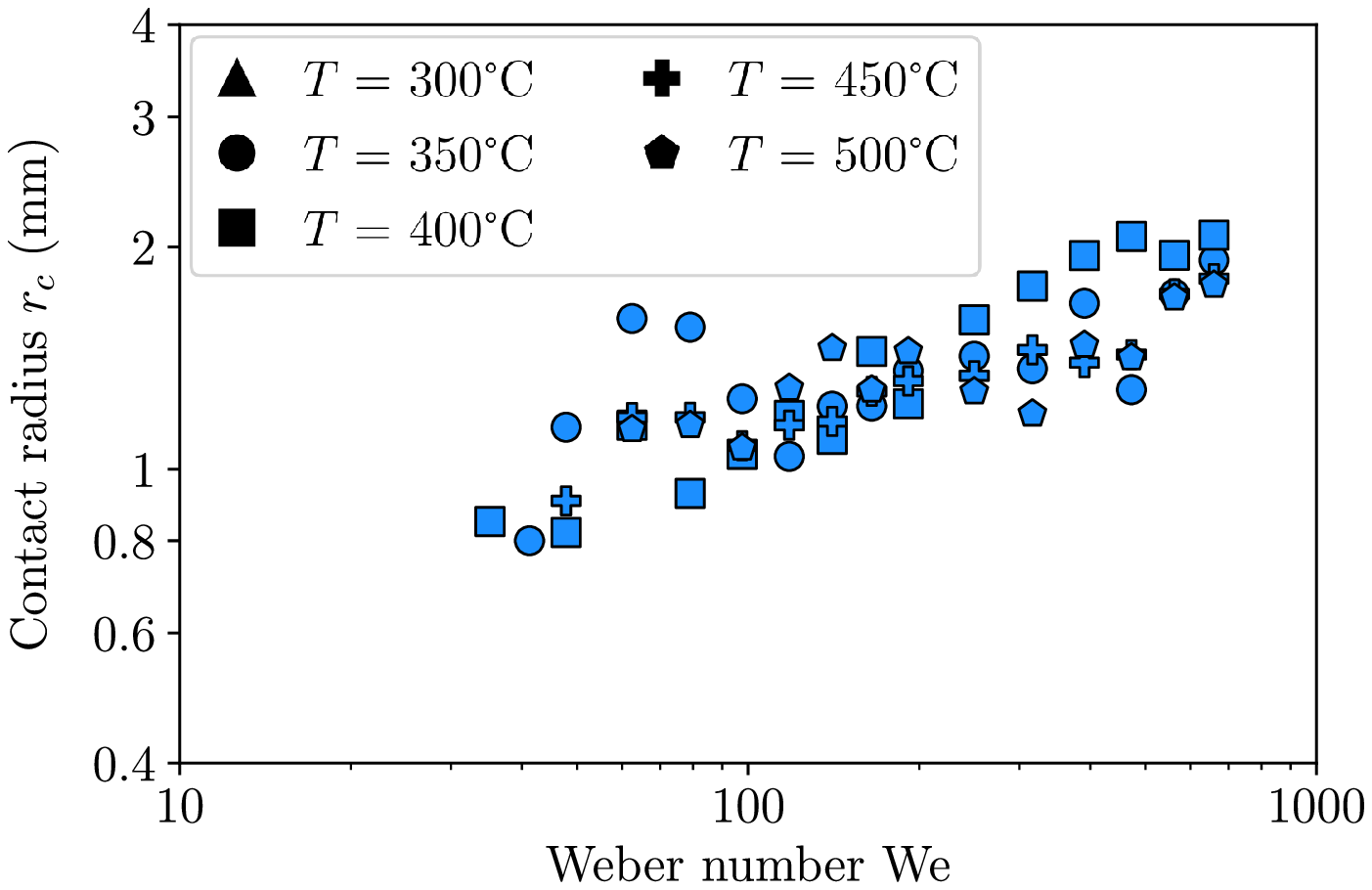}
        \subcaption{Fixed jet radius $a=195 \ \si{\micro\metre}$}
        \label{subfig:d_contact_fixed_needle}
    \end{subfigure}\hfill
    \begin{subfigure}[t]{.5\linewidth}
        \centering
        \vskip 0pt
        \includegraphics[width=\textwidth]{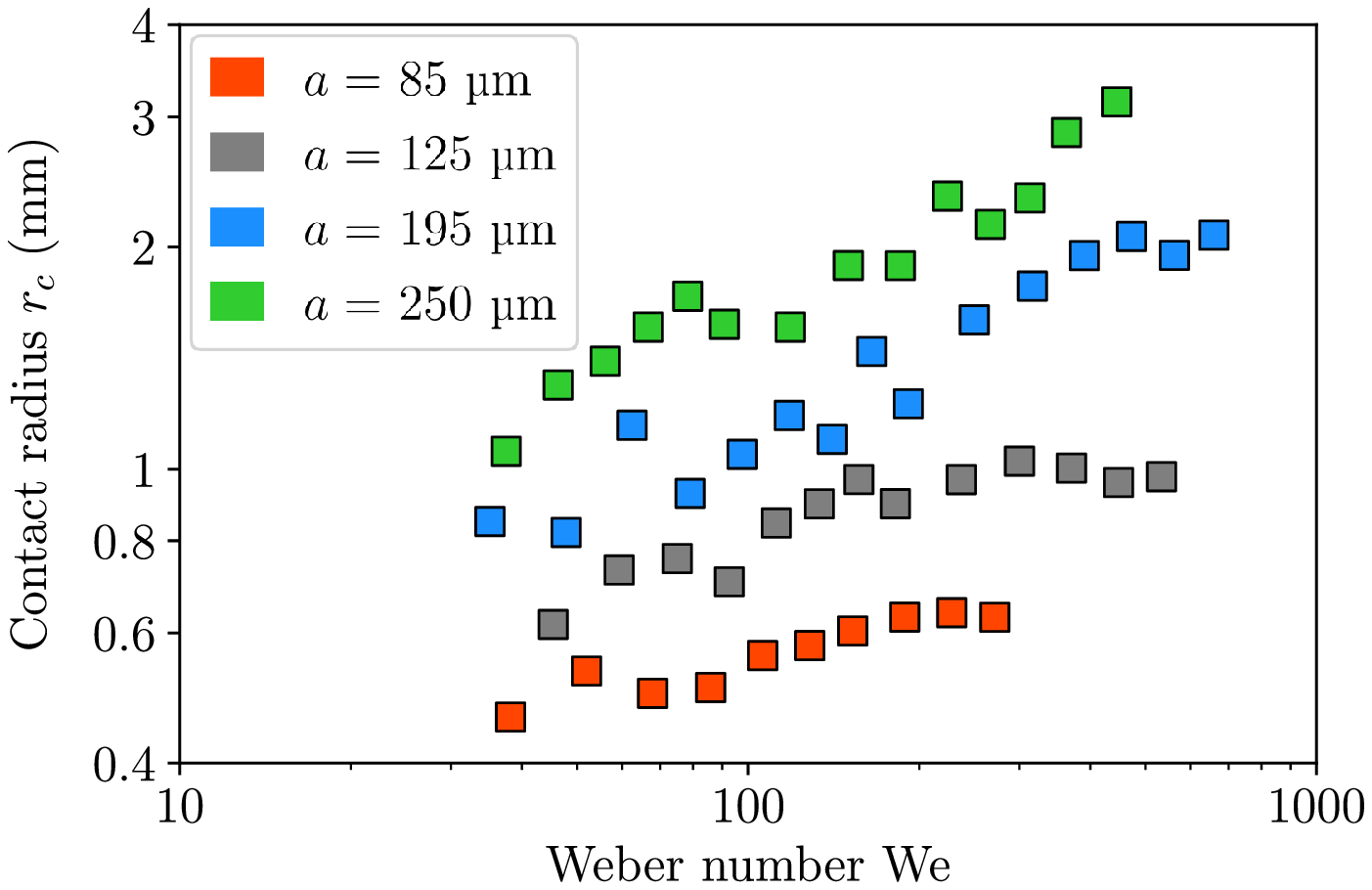}
        \subcaption{Fixed surface temperature $T=400 \si{\degC}$}
        \label{subfig:d_contact_fixed_temp}
    \end{subfigure}
    \caption{\justifying The radius $r_c$ of the contact area between the water and the heated surface plotted against the Weber number for either a fixed (a) jet radius $a=195 \ \si{\micro\metre}$ or a fixed (b) surface temperature $T=400 \si{\degreeCelsius}$. The data indicates a power-law relationship between $r_c$ and $\We$, with a clear dependency on the jet radius.}
    \label{fig:d_contact}
\end{figure}

Experimental results show that both $\theta$ and $r_c$ remain constant over time, provided that the jet flow rate remains unchanged. This indicates that our experiments are indeed conducted in a stationary regime. Regarding the variations in $\theta$, the results are illustrated in Fig. \ref{fig:theta}, confirming a decrease in the ejection angle as the Weber number increases. All experimental points collapse on the same curve irrespective of the surface temperature $T$ and jet radius $a$, strongly suggesting that the ejection angle $\theta$ is not influenced by these two parameters. As for the radius $r_c$ of the contact area, we noticed its increase with $\We$, a trend that aligns with intuition given the rise in inertia with $\We$. We observed a power law dependency for $r_c$ with respect to We, and, similarly to $\theta$, the temperature $T$ of the disk also has no impact on $r_c$ (Fig. \ref{subfig:d_contact_fixed_needle}). However, this similarity ends with the clear influence of the jet radius $a$ over $r_c$ (Fig. \ref{subfig:d_contact_fixed_temp}) as $r_c$ increases with $a$. Measurement uncertainties are estimated to be at most 10\% for $\theta$ and $r_c$. In the following section, we will discuss these results.

\section{Discussion}

\begin{figure}[H]
    \centering
    \includegraphics[width=.35\linewidth]{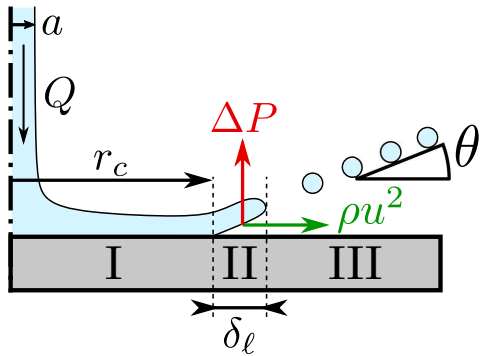}
    \caption{\justifying Division of the study into 3 areas: the \textit{wet area} (zone I) where the water is in contact with the heated surface, the \textit{launching area} (zone II) where a vapor layer lifts the water off the surface and the \textit{fragmentation area} (zone III) where the liquid sheet breaks into small droplets.}
    \label{fig:zones}
\end{figure}

The upcoming sections will introduce simple models founded on three distinctive regions that emerge experimentally and theoretically \cite{agrawalSurfaceQuenchingJet2019} with the distance from the point of impact of the jet on the hot plate. We refer to them as labelled in Fig. \ref{fig:zones} as zone (I), (II) and (III). Hereafter, we first describe their characteristics before providing models for each of them in the subsequent sections.

Zone (I) corresponds to the area close to the impact point and is called \textit{wet area}. The pressure imposed by the jet prevents the vapor layer of the Leidenfrost effect from forming. Therefore, the water in this area is in direct contact with the heated surface. The radius of this area is $r_c$.

Zone (II), the \textit{launching area}, delineates the area where the liquid layer separates from the heated surface, indicating the formation of an underlying vapor layer. This region also influences the angle of droplet ejection, which is one of the distinguishing features of zone (III). This last zone, called the \textit{fragmentation area}, witnesses the fragmentation of the layer into tiny droplets similar to a Savart liquid sheet. The droplets are ejected in all directions at an angle $\theta$ relative to the horizontal.

\subsection{Zone I: the wet area}
\label{subsect:zone1}

As previously mentioned, in this region, the water maintains contact with the disk even at plate temperatures reaching up to 500\si{\degC}, which is below the minimum film boiling temperature of 840°C predicted by Karwa \textit{et al.} \cite{karwaHydrodynamicModelSubcooled2011}. The contact area being characterized by $r_c$, we shall therefore endeavour to obtain an associated scaling law with respect to experimental parameters. To this end, we consider $r_c$ to be the distance the water must travel away from the jet to be heated from its initial temperature of 20°C up to 100°C.
We define the time $\tau$ needed for the water to travel from the impact point of the jet to the edge of the wet area, \textit{i.e.} $r_c$. The characteristic velocity in this simplified model is defined as $v=Q/\pi a^2$, assuming a fully inertial regime, leading to the expression for the characteristic time $\tau$ as:
\begin{equation}
    \tau = \frac{r_c}{v} \sim \frac{\pi a^2}{Q}r_c.
    \label{eq:tau}
\end{equation}
During this period, the water receives a heat flux $\varphi_\mathrm{plate}$ from the hot plate which can be expressed as
\begin{equation}
    \varphi_\mathrm{plate}=h_\mathrm{heat} \Delta T_\mathrm{plate}
    \label{eq:phi}
\end{equation}
where $h_\mathrm{heat}$ is the heat transfer coefficient, expressed in \si{\watt\per\square\metre\per\kelvin}, and $\Delta T_\mathrm{plate}=T_\mathrm{plate}-T_\mathrm{amb}$ denotes the temperature difference between the plate ($T_\mathrm{plate}$) and the water ($T_\mathrm{amb}=20\si{\degC}$), which corresponds to the ambient temperature. Assuming that the heat transfer occurs only through convection, we can estimate $h_\mathrm{heat}$ using a conductive model, yielding
\begin{equation}
    h_\mathrm{heat}\sim \frac{\lambda}{e}
    \label{eq:h}
\end{equation}
where $\lambda=0.6$ \si{\watt\per\metre\per\kelvin} is the thermal conductivity of liquid water and $e$ is the average thickness of the water layer. Assuming $e$ remains constant across the entire wet area, the volume $V$ of fluid that has flowed during the time $\tau$ is given by
\begin{equation}
    V= Q \tau \quad \text{and} \quad V= \pi r_c^2 e.
\end{equation}
Using equation (\ref{eq:tau}), these two expressions yield
\begin{equation}
    e\sim\frac{a^2}{r_c}.
    \label{eq:e}
\end{equation}
Finally, we calculate the energy absorbed by the water from the plate as it travels from the point of impact of the jet to the end of the wetted region, given by
\begin{equation}
    \mathcal{E}_\mathrm{rec}=\varphi_\mathrm{plate} \tau S
\end{equation}
where $S$ is the surface of the wetted area. Using equations (\ref{eq:tau}), (\ref{eq:phi}), (\ref{eq:h}) and (\ref{eq:e}), one obtains
\begin{equation}
    \mathcal{E}_\mathrm{rec}\sim\frac{\pi \lambda r_c^2}{Q} S \Delta T_\mathrm{plate}.
    \label{eq:e_rec}
\end{equation}

To establish a scaling law for $r_c$, we presume that the energy $\mathcal{E}_\mathrm{rec}$ transferred to the water is offset by the energy required to elevate its temperature from its initial ambient temperature $T_\mathrm{amb}=20\si{\degreeCelsius}$ to its boiling temperature $T_\mathrm{boil}=100\si{\degreeCelsius}$, facilitating the boiling of water and the formation of a vapor layer in the launch area. An estimate of this energy can be obtained by:
\begin{equation}
    \mathcal{E}_\mathrm{boil}=\rho C_p V \Delta T_\mathrm{water}
    \label{eq:e_boil_0}
\end{equation}
where $C_p=4187$ \si{\joule\per\kilo\gram\per\kelvin} is the specific heat of water, $V$ is the volume of water that is boiled and $\Delta T_\mathrm{water}=T_\mathrm{boil}-T_\mathrm{amb}$. The heated volume of water $V$ can be estimated by the expression $V=Se$. Recalling the expression for $e$ given by equation (\ref{eq:e}), equation (\ref{eq:e_boil_0}) reads
\begin{equation}
    \mathcal{E}_\mathrm{boil}=\rho C_p \frac{a^2}{r_c} S\Delta T_\mathrm{water}.
    \label{eq:e_boil}
\end{equation}

\begin{figure}[H]
    \centering
    \includegraphics[width=.5\linewidth]{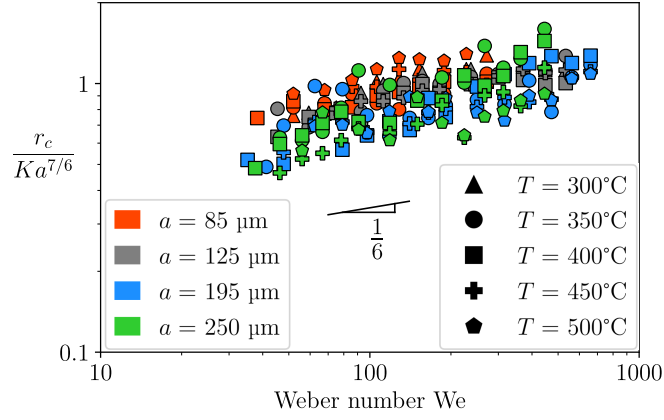}
    \caption{Scaling of $r_c$ with respect to $\We$ obtained via equation (\ref{eq:lc_scaling}), where $K$ is a constant derived from equation (\ref{eq:lc_scaling}).}
    \label{fig:lc_model_scaling}
\end{figure}

We can now derive an expression for $r_c$ by equating equations (\ref{eq:e_rec}) and (\ref{eq:e_boil}), resulting in:
\begin{equation}
    \mathcal{E}_\mathrm{boil}= \mathcal{E}_\mathrm{rec}\Rightarrow r_c \sim Q^{1/3}a^{2/3}\left(\frac{\rho C_p \Delta T_\mathrm{water}}{\pi \lambda \Delta T_\mathrm{plate}}\right)^{1/3}. 
    \label{eq:lc_wrt_Q}
\end{equation}
We make one final assumption in this simplified model, regarding $\Delta T_\mathrm{plate}$. Data presented in Fig.~\ref{subfig:d_contact_fixed_needle} indicates that $r_c$ does not depend on the temperature of the heated surface. Therefore, we hypothesize that the localized surface beneath the wet area (i.e., directly beneath the jet) is sufficiently cooled to maintain a temperature equal to that of boiling water, i.e., $100\si{\degC}$, while every other point on the heated surface exceeds the prescribed temperature. One can note that this assumption is consistent with the expression of the heat transfer coefficient $h_\mathrm{heat}$ between the plate and the water as we use the thermal conductivity $\lambda$ of liquid water. Therefore, we write $\Delta T_\mathrm{plate}=\Delta T_\mathrm{water}=80 ^{\circ} $ \si{\kelvin}. Finally, equation (\ref{eq:lc_wrt_Q}) is re-expressed in terms of $\We$ by substituting $Q$ using equation (\ref{eq:weber}):
\begin{equation}
    r_c \sim \We^{1/6}a^{7/6}\left(\frac{C_p}{\lambda}\sqrt{\frac{\gamma\rho}{2}}\right)^{1/3}.
    \label{eq:lc}
\end{equation}
As a result, we can establish the following scaling law:
\begin{equation}
    \frac{r_c}{Ka^{7/6}}\propto \We^{1/6}, 
    \label{eq:lc_scaling}
\end{equation}
where $K=\left(\frac{C_p}{\lambda}\sqrt{\frac{\gamma\rho}{2}}\right)^{1/3}$. This scaling law is compared to our experimental data in Figure \ref{fig:lc_model_scaling}. Our measurements collapsed relatively well and the 1/6 exponent corresponds closely with our data. Furthermore, equation (\ref{eq:lc}) allows us to estimate $r_c$. For instance, with $a=125$ \si{\micro\metre} and $\We=60$, we obtain $r_c \sim 2$ \si{\milli\metre}, which is within the expected range compared to Fig. \ref{fig:lc_model_scaling}. Hence, our simple model yields reasonably accurate results.

\subsection{Zone II: the launching area}
\label{subsect:zone2}

In this area, the water layer detaches from the heated surface as a vapor layer forms beneath it. The objective of this section is to characterize the transition from the growing drop regime to the droplet ejection regime. Furthermore, we aim to  elucidate the observed value of the Weber number of approximately 35 for the jet at this transition. To achieve this, we assume the water temperature in the launching area is around 100\si{\degC}, and we introduce a local Weber number $\Weloc$, defined at the distance $r_c$ from the jet: 
\begin{equation}
    \Weloc=\frac{\rho_{100}u_\mathrm{loc}^2 h_\mathrm{loc}}{\gamma_{100}}=\frac{\rho_{100}u^2(r_c) h(r_c)}{\gamma_{100}},
    \label{eq:weloc}
\end{equation}
where $\rho_{100}=960$~\si{\kilo\gram\per\cubic\metre} is the density of water at 100\si{\degC}, $\gamma_{100}=59$~\si{\milli\newton\per\metre} is the surface tension of water at 100\si{\degC} and $u(r_c)$ and $h(r_c)$ are the local speed and the depth, respectively, of the flow in the launching area. As a Weber number compares the effect of inertia over the effect of surface tension, the two observed regimes can be rationalized as follows. For $\Weloc<1$ the droplet lacks sufficient inertia to overcome surface tension effects, causing the drop to grow as it is supplied by the jet. Conversely, for $\Weloc>1$ surface tension effects are insufficient to counteract inertia, resulting in fragmentation of the water-air interface into small droplets.

Previous studies in the literature have documented instances of instabilities in radially expanding liquid sheets \cite{linAbsoluteConvectiveInstability2003,maynesFreesurfaceLiquidJet2011}. These investigations revealed that the local Weber number approached unity at the onset of instability. Consequently, we aim to derive an expression for $\Weloc$ and determine the Weber number of the jet at the transition point, namely when $\Weloc = 1$.
To achieve this, we employ the expressions for $u(r_c)$ and $h(r_c)$ as given by Watson's equations \cite{watsonRadialSpreadLiquid1964}, which describe the radial spread of a liquid jet over a horizontal surface. Consequently, the height of the water layer at $r_c$ can be expressed as \cite{watsonRadialSpreadLiquid1964}:
\begin{equation}
   h(r_c) = \frac{2\pi^2\nu_{100}}{3\sqrt{3}}\frac{r_c^3+\ell^3}{Qr_c},
    \label{eq:watson_height_lc_with_ell}
\end{equation}
where $\ell=0.567 a \Re^{1/3}$ represents a length derived from computation, $\Re=Q/a\nu_{100}$ denotes the Reynolds number of the flow, and $\nu_{100}=0.3\cdot10^{-6}$ \si{\square\metre\per\second} stands for the kinematic viscosity of liquid water at 100\si{\degC}. One can note that in our case, $\ell \sim 2\cdot 10^{-3}$ \si{\metre} and therefore is not negligible at all compared to $r_c \sim 10^{-3}$ \si{\metre}. The local flow velocity is then determined by the conservation of mass:
\begin{equation}
    u(r_c)=\frac{Q}{2\pi r_c h(r_c)}=\frac{3\sqrt{3}}{4\pi^3\nu_{100}}\frac{Q^2}{r_c^3+\ell^3}.
    \label{eq:watson_speed_lc}
\end{equation}
Equation (\ref{eq:weloc}) can then be rewritten as
\begin{equation}
        \We_\mathrm{loc}
        =\frac{3\sqrt{3}}{8\pi^4}\frac{\rho_{100}}{\gamma_{100}\nu_{100}} \frac{Q^3}{r_c(r_c^3+\ell^3)}.
\end{equation}
By replacing $r_c$ by its expression in eq.~(\ref{eq:lc_wrt_Q}), one obtains 
\begin{equation}
    \Weloc=\frac{3\sqrt{3}}{8\pi^4}\frac{\rho_{100}}{\gamma_{100}\nu_{100}}\frac{\left(\dfrac{\pi\lambda}{\rho C_p}\right)^{1/3} }{\dfrac{\rho C_p}{\pi\lambda}+\dfrac{0.567^3}{\nu_{100}}}Q^{5/3}a^{-8/3}.
    \label{eq:weloc_wrt_Q}
\end{equation}

We can compute $Q$ with respect to $\Weloc$ and $a$ using equation (\ref{eq:weloc_wrt_Q}), and then use equation (\ref{eq:weber}) to express the Weber number of the jet in relation to $\Weloc$ as:
\begin{equation}
        \We=\left[\frac{8\pi^4}{3\sqrt{3}}\frac{\gamma_{100}\nu_{100}}{\rho_{100}}\left(\frac{\rho C_p}{\pi\lambda}\right)^{1/3}\left(\frac{\rho C_p}{\pi\lambda}+\frac{0.567^3}{\nu_{100}}\right)\right]^{6/5}\frac{2\rho}{\pi\gamma}a^{1/5}\We_\mathrm{loc}^{6/5}.
        \label{eq:we_wrt_weloc}
\end{equation}

Given that \( \mathrm{We}_{\mathrm{loc,trans}} = 1 \) at the transition point, one can infer from this equation the corresponding value of $\We_\mathrm{trans}$.
One can note a very slight dependency on $a$, which is not experimentally observable given the range of jet radii we explored. The absence of the hot plate temperature on $\We_\mathrm{trans}$ is also consistent with our observations. For a needle of radius $a = 195$ \si{\micro\metre}, equation (\ref{eq:we_wrt_weloc}) yields
\begin{equation}
    \We_\mathrm{trans,model} \sim 515.
\end{equation}
This value far exceeds our experimental observations. However, if we use experimental values for $r_c$ ($\sim 0.8$ mm) instead of relation (\ref{eq:lc_wrt_Q}), we recover
\begin{equation}
    \We_\mathrm{trans,meas} \sim 20, 
\end{equation}
which is actually quite close to the expected value of 35.

\subsection{Zone III: the fragmentation area}
\label{subsect:zone3}

In this zone, the liquid sheet, upon detachment from the heated surface in the launching region, disintegrates into small droplets. This region is characterized by an angle of ejection, denoted as \( \theta \), which specifies the trajectory of the droplets relative to the horizontal plane.

To elucidate the ejection angle, one may revisit the launching area and examine the influence of the following two forces  acting on a droplet (refer to Fig. \ref{fig:zones}). The first force, stemming from the vaporization of water into the vapor layer in the launching area, induces an overpressure, lifting the water layer from the heated surface and resulting in an upward force denoted as $\Fup$. As the thickness of the liquid sheet is a few dozen micrometres, this force is likely to have a substantive impact on the sheet. The second force, attributed to the inertia of the water, creates a radial pressure that propels the water outward from the jet, thereby generating a radial outward force denoted as \( \Frad \). The interplay of these two forces leads to a diagonal trajectory and an ejection angle \( \theta \) for the droplets, which can be formulated as
\begin{equation}
    \tan(\theta)=\frac{\Fup}{\Frad}
    \label{eq:def_tan_theta}
\end{equation}

\begin{figure}[H]
    \centering
    \includegraphics[width=.5\linewidth]{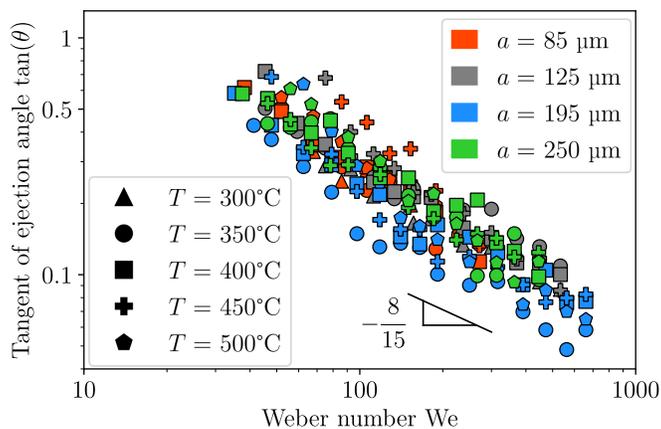}
    \caption{\justifying Scaling of $\tan(\theta)$ with respect to $\We$ obtained via equation (\ref{eq:theta_scaling}). The predicted $-8/15$ exponent for $\We$ is in a relatively good agreement with our experimental measurements.}
    \label{fig:theta_scaling} 
\end{figure}

\noindent We can estimate $\Frad$ by considering that an outward radial pressure driven by inertia is applied on the inner vertical section of the water layer in the launching area. Therefore,
\begin{equation}
    \Frad = \frac{1}{2}\rho_{100} u(r_c)^2 \times 2\pi r_c h(r_c).
    \label{eq:def_Frad}
\end{equation}
By applying the conservation of mass, $Q=2\pi r_c h(r_c) u(r_c)$, and substituting the expression of $u(r_c)$ from Eq.~(\ref{eq:watson_speed_lc}) into equation (\ref{eq:def_Frad}), we derive:
\begin{equation}
    \Frad = \frac{3\sqrt{3}}{8\pi^3}\frac{\rho_{100}}{\nu_{100}}\frac{Q^3}{r_c^3+\ell^3}
    \label{eq:frad}
\end{equation}
For the upward force \( \Fup \), we can reasonably assume that an overpressure \( \Delta P \) acts on the bottom surface of the water layer in the launching area. By defining the width of the launching area as \( \delta_\ell \) (see Fig. \ref{fig:zones}), the force \( \Fup \) can be expressed as:
\begin{equation}
    \Fup=\Delta P \pi\left[(r_c+\dl)^2-r_c^2\right]
    \label{eq:def_Fup}
\end{equation}
One can approximate \( \Delta P \) using an analogy with Celestini's study \cite{celestiniTakeSmallLeidenfrost2012}, which investigates the take-off of Leidenfrost drops and provides the following expression for the overpressure within the vapor layer:
\begin{equation}
    \Delta P \sim L^2 \frac{\nu_v\lambda_v \Delta T_v}{h_v^4 \mathcal{L}},
    \label{eq:deltaP}
\end{equation}
where $L$ is a characteristic length of the vapor layer, $\nu_v= 2\cdot10^{-5}$ \si{\square\metre\per\second} is the viscosity of water vapor at 100\si{\degC} and ${\lambda_v=0.03}$~\si{\watt\per\metre\per\kelvin} its thermal conductivity at 100\si{\degC}, $\mathcal{L}=2.26\cdot10^6$ \si{\joule\per\kilogram} is the latent heat of vaporization of water, $\Delta T_v=T-T_\mathrm{boil}$ is the difference between the temperature of the hot plate in the launching area and the temperature at the top of the water layer (100\si{\degC}) and $h_v$ is the thickness of the vapor layer in the launching area. As $L^2$ represents a characteristic surface area for the vapor layer, we assume it can be substituted by the surface area of the vapor layer in the launching area, given by \( \pi\left[(r_c+\delta_\ell)^2-r_c^2\right] \). The upward force can thus be expressed as
\begin{equation}
    \Fup\sim\pi^2\left[(r_c+\dl)^2-r_c^2\right]^2\frac{\nu_v \lambda_v \Delta T_v}{h_v^4 \mathcal{L}}
    \label{eq:fup}
\end{equation}
Combining equations (\ref{eq:def_tan_theta}), (\ref{eq:frad}) and (\ref{eq:fup}) yields 
\begin{equation}
    \tan(\theta)\sim \frac{8\pi^5}{3\sqrt{3}}\frac{\nu_{100}\nu_v\lambda_v\Delta T_v}{\rho_{100} h_v^4\mathcal{L}}\frac{\dl^2\left(2r_c+\dl\right)^2(r_c^3+\ell^3)}{Q^3}
    \label{eq:tan_theta}
\end{equation}
Experimental observations indicate that $\dl\ll r_c$. Additionally, we can estimate the height of the vapor layer by writing $\tan(\theta)\sim h_v/\dl$. Consequently, equation (\ref{eq:tan_theta}) becomes
\begin{equation}
    \tan^5(\theta)\sim \frac{32\pi^5}{3\sqrt{3}}\frac{\nu_{100}\nu_v\lambda_v\Delta T_v}{\rho_{100} \mathcal{L}}\frac{r_c^2(r_c^3+\ell^3)}{\dl^2 Q^3}
    \label{eq:tan_theta_simpl}
\end{equation}
We can derive from this equation an approximate value of $\tan(\theta)$. We estimate \( \delta_\ell \) to be around 0.1 mm, a reasonable approximation based on our experimental findings. Considering \( a = 195  \)~\si{\micro\metre}, $\We \sim 50$, and utilizing equations (\ref{eq:weber}) and~(\ref{eq:lc}) as well as the fact that $\ell=0.567\nu_{100}^{-1/3}a^{2/3}Q^{1/3}$, we obtain \( \tan(\theta) \approx 0.52\), which aligns with the expected order of magnitude when compared with our measured values presented in Fig. \ref{fig:theta_scaling}.
However, as $\delta_\ell$ is not accessible experimentally, we need to obtain a relation for $\delta_\ell$. To do so, we assume that only inertia has an impact on the water layer in the launching area. The water layer is therefore equivalent to a Savart sheet in this area, and as $\We<1000$ in our experiments, we can write $\dl \propto \We$ \cite{clanetLifeSmoothLiquid2002}. Using this estimation along with equations (\ref{eq:weber}) and (\ref{eq:lc}), one obtains:
\begin{equation}
    \tan(\theta)\propto a^{4/15} \We^{-8/15} \Delta T_v^{1/5}
    \label{eq:theta_scaling}
\end{equation}
This scaling law is compared to experimental data in Figure \ref{fig:theta_scaling}. The experimental results suggest a power dependency of \( \We\) with an exponent of \(-8/15\), akin to the model, particularly when examining data for a constant jet diameter. Nevertheless, the collective dataset appears to exhibit a slight deviation from the anticipated power of \(-8/15\). This discrepancy can be attributed to a nuanced reliance on both \(a\) and \(T\) (via \(\Delta T_v\)) in the model, which is not accounted for in the plotted data. Our proposed explanation for the observed weak dependency is that the plate is locally cooled to such an extent by the jet that the initial temperature near the jet impact point becomes irrelevant.

Having established the scaling laws for different zones, we now turn our attention to the energy contained in droplets during atomization. This will enable us to estimate the efficiency of our system in breaking up a liquid sheet into droplets and to compare our results with similar data in the literature that does not involve heating, such as the Savart sheet reported by Clanet \textit{et al.}~\cite{clanetLifeSmoothLiquid2002}. To achieve this, we need to measure the size and velocity of the ejected droplets, which is the focus of the next section.

\section{Energy of atomization}
\begin{figure}[H]
    \centering
    \begin{subfigure}[b]{.5\linewidth}
        \centering
        \includegraphics[height=.255\textheight]{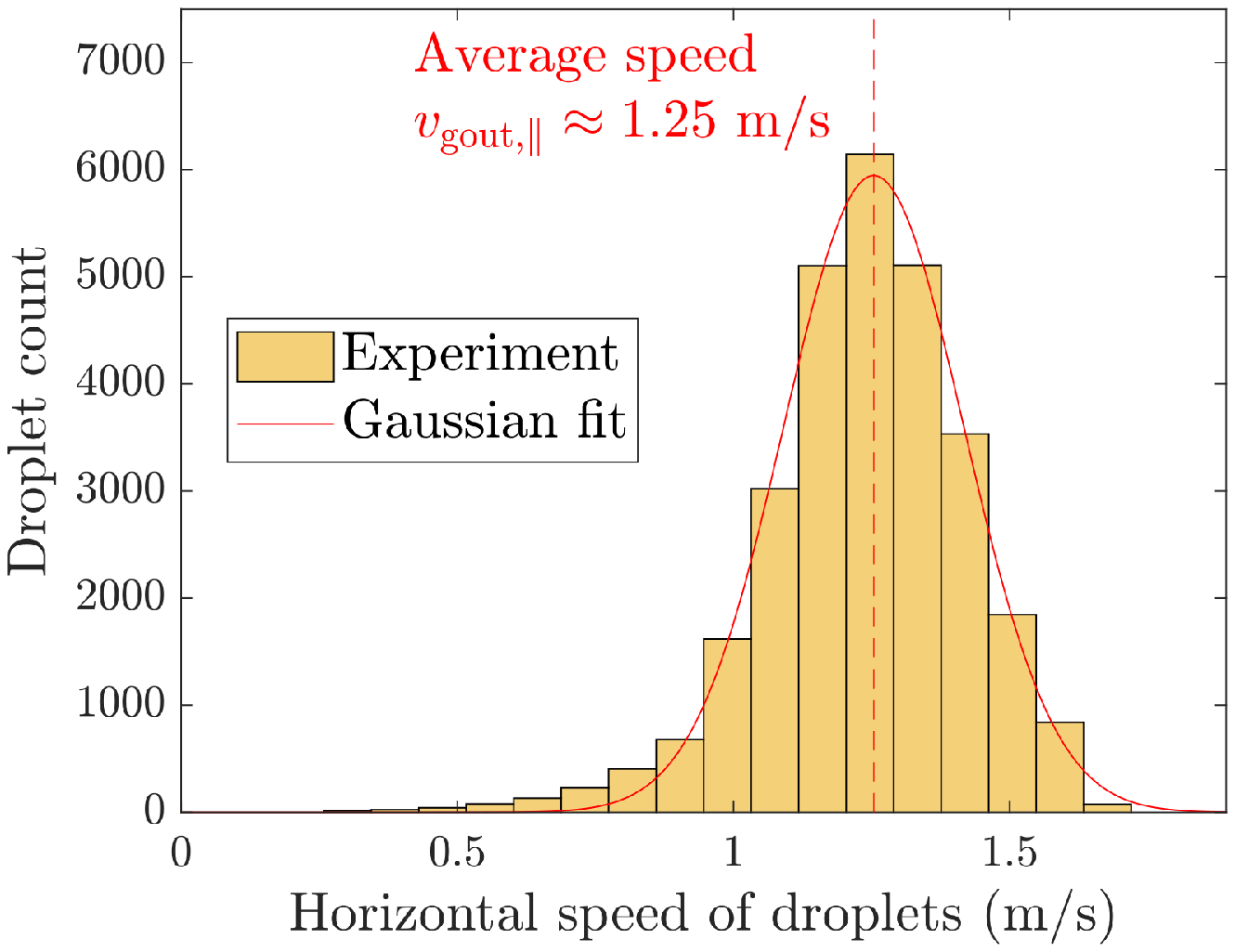}
        \vskip 6pt
        \subcaption{Horizontal speed distribution.}
        \label{subfig:distrib_vit_droplet}
    \end{subfigure}\hfill
    \begin{subfigure}[b]{.5\linewidth}
        \centering
        \vskip 0pt
        \includegraphics[width=\textwidth]{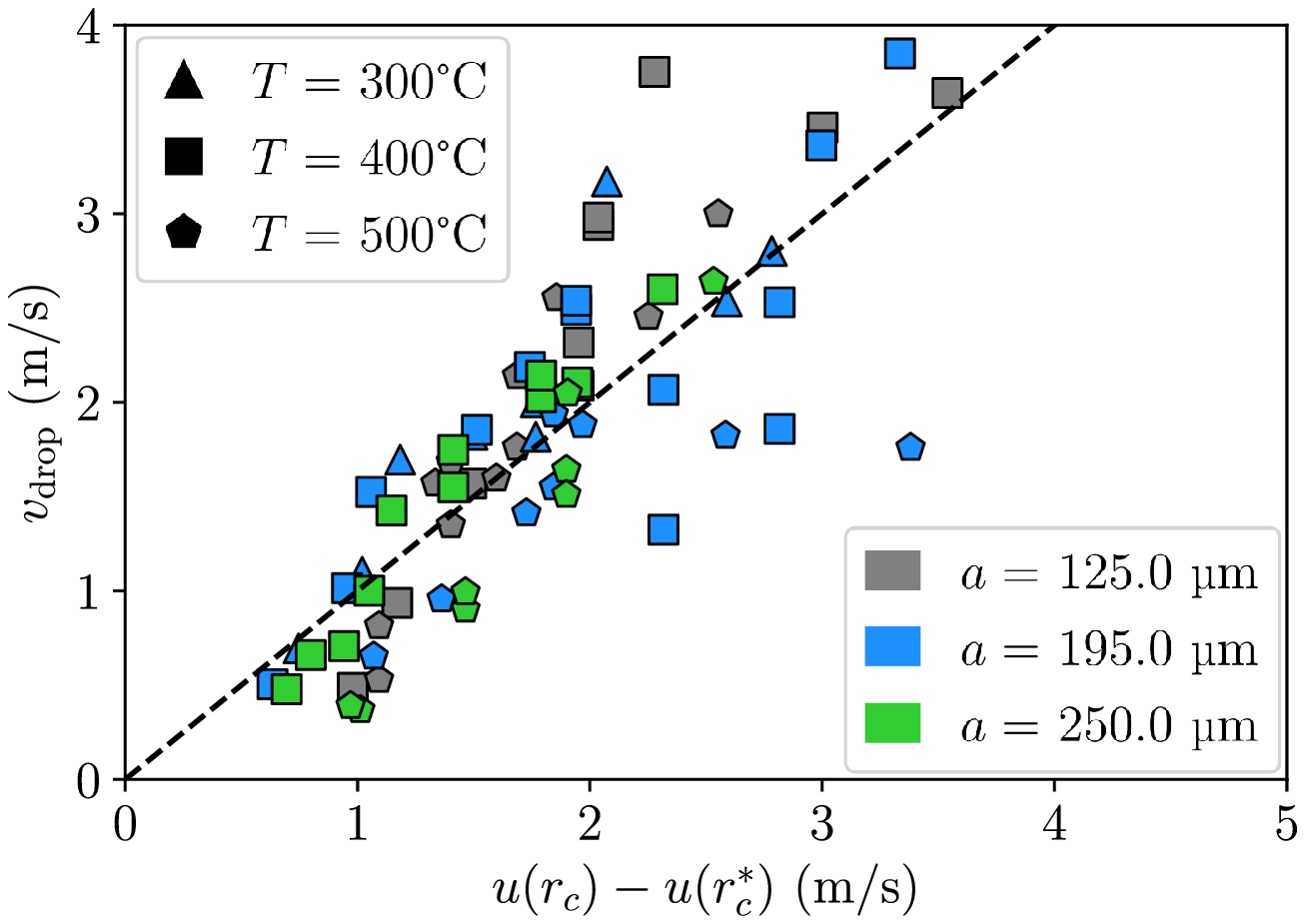}
        \subcaption{Average droplet speed.}
        \label{subfig:vit_droplets}
    \end{subfigure}
    \caption{\justifying Horizontal speed of ejected droplets for different jet radii $a$ and plate temperatures $T$. (a) Example of an observed distribution for the measured horizontal speed of the droplets for $a=195$ \si{\micro\metre}, $T=300$\si{\degC} and $\We = 79$. (b) Average droplet speed $v_\mathrm{drop}$ with respect to $u(r_c)-u(r_c^*)$, where $u(r_c^*)$ is the speed of the flow in the liquid sheet when $\Weloc = 1$. A linear dependence can be observed between the two parameters. The black dotted line indicates a slope of 1.}
    \label{fig:vit_droplets}
\end{figure}

In this section, we aim to characterize the droplets emitted from the fragmentation area. Specifically, we explore how experimental parameters impact their properties, including velocities and size. A Matlab code employing a tracking algorithm \cite{crockerMethodsDigitalVideo1996} has been developed and implemented to extract the trajectories and horizontal velocities of the droplets from the recorded videos. Figure~\ref{subfig:distrib_vit_droplet} shows a typical velocity distribution in the experiments. Fitting these distributions with a Gaussian curve allows us to determine the average horizontal velocity for the droplets. Given that we also have information about the ejection angle of the droplets (see Fig. \ref{fig:theta}), we are able to calculate the average magnitude of the velocity of the droplets, denoted as $v_\mathrm{drop}$. 

Figure \ref{subfig:vit_droplets} illustrates $v_\mathrm{drop}$ plotted against  $u(r_c)-u(r_c^*)$, where $u(r_c^*)$ represents how much speed the droplets lost in order to overcome surface tension. To compute $u(r_c^*)$, we set $\Weloc = 1$ in equation (\ref{eq:weloc}). We then approximate the radius of contact $r_c^*$ and the flowrate $Q^*$ at the transition between regimes (i) and (ii) by using our measurements for the lowest Weber number at which regime (ii) is observed in figure \ref{subfig:d_contact_fixed_temp}. Our data then show that $v_\mathrm{drop}$ matches quite well $u(r_c)-u(r_c^*)$, indicating that the inertia of the droplets is indeed what remains of the inertia of the liquid sheet after surface tension is overcome. 
One can notice that neither $T$ nor $a$ affect $v_\mathrm{drop}$.

\begin{figure}[H]
    \centering
    \begin{subfigure}[t]{0.49\textwidth}
        \centering
        \vskip 0pt
        \includegraphics[width=\textwidth]{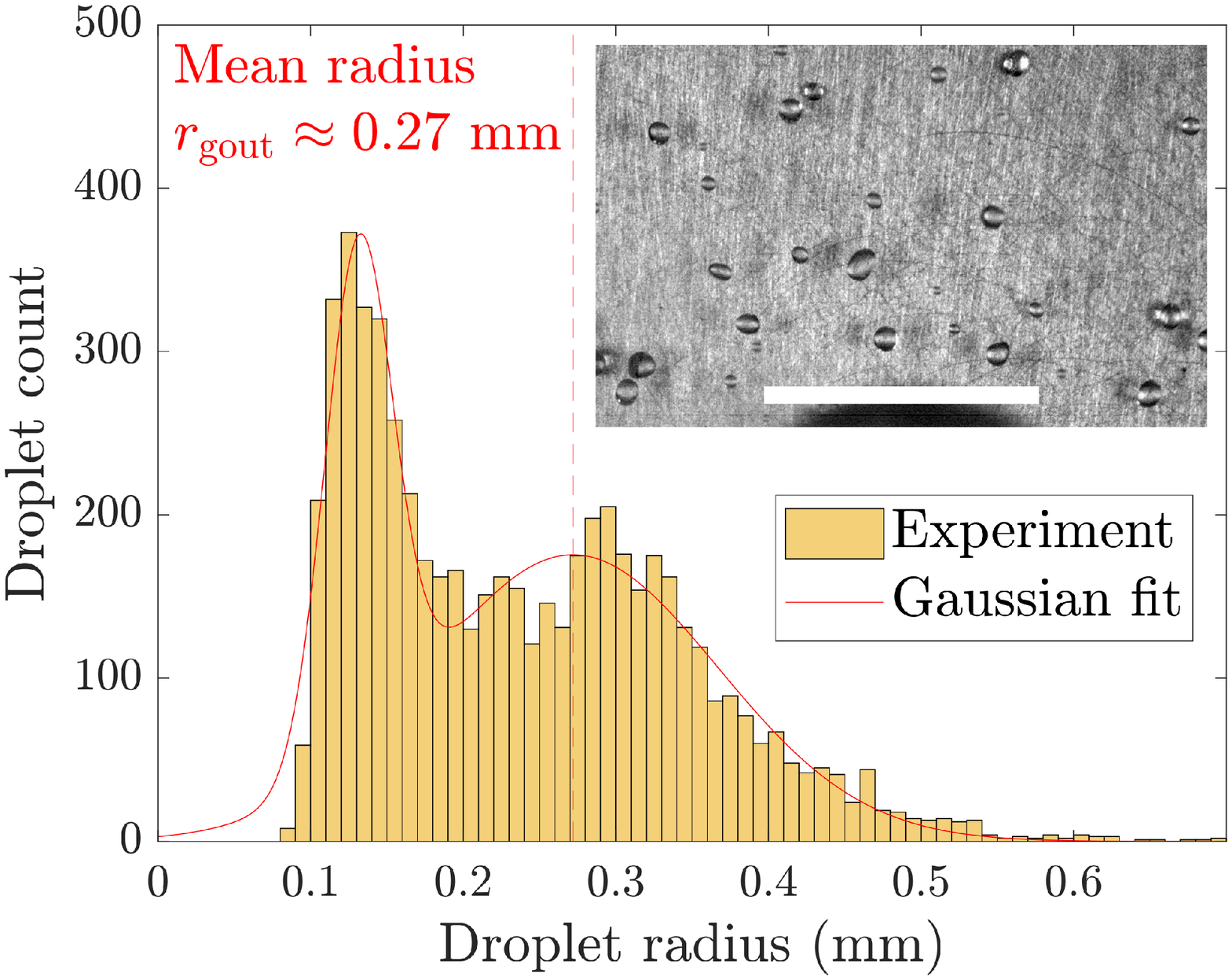}
        \subcaption{$\We \simeq 38$}
        \label{subfig:distrib_rayon_bidisperse}
    \end{subfigure}\hfill
    \begin{subfigure}[t]{0.49\textwidth}
        \centering
        \vskip 0pt
        \includegraphics[width=\textwidth]{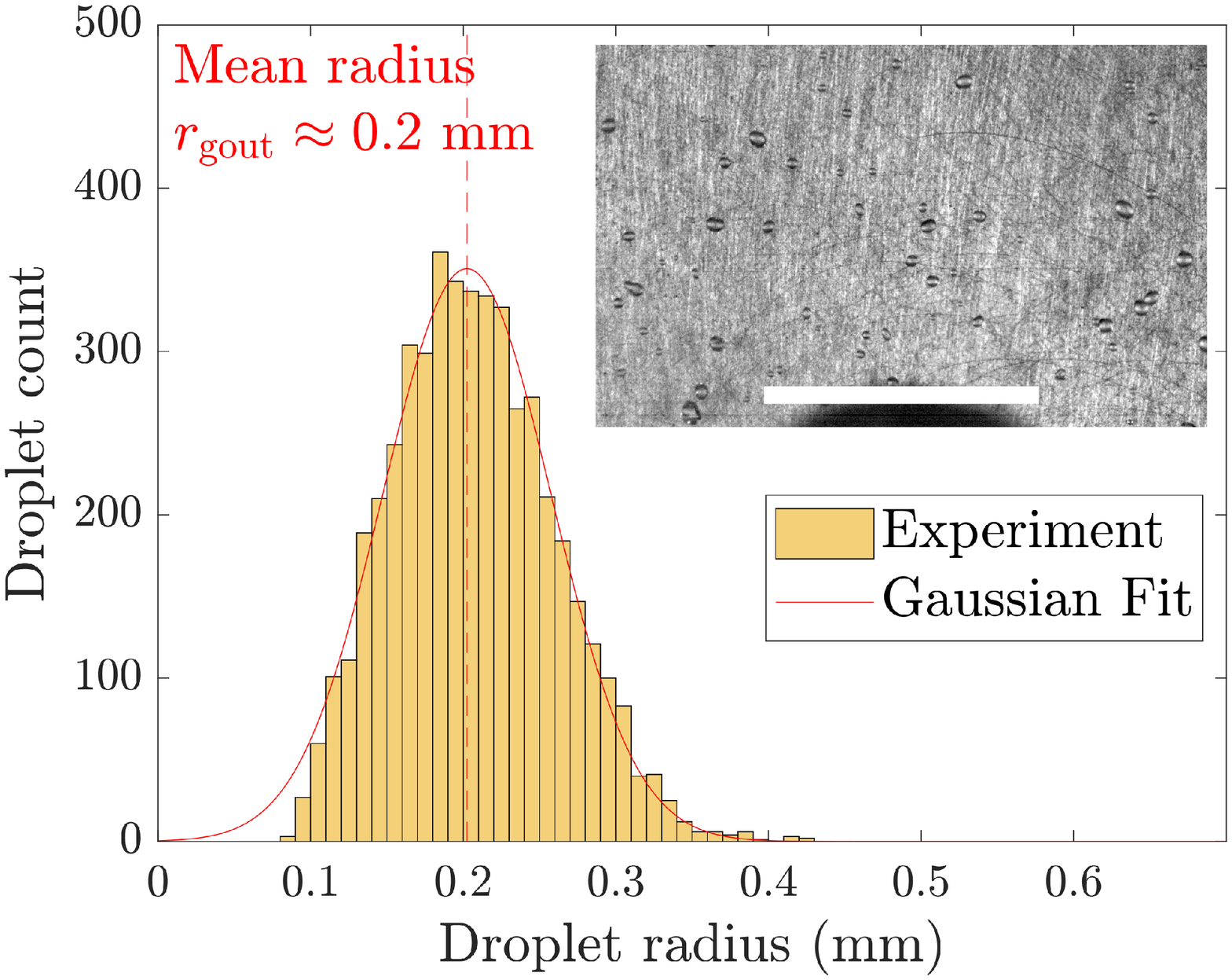}
        \subcaption{$\We \simeq 68$}
        \label{subfig:distrib_rayon_monodisperse}
    \end{subfigure}
    \caption{\justifying Radius distribution of the ejected droplets when a jet of radius $a=250$ \si{\micro\metre} impacts the Plexiglas plate heated at $T=400$\si{\degC} for two different Weber numbers. (a) For $\We\simeq 38$, we observe a bidisperse distribution in the radii of the ejected droplets. (b) For $\We\simeq 68$, the distribution becomes monodisperse. Insets show a top-view of the ejected droplets. White scale bar is 1 cm.}
    \label{fig:distrib_rayon}
\end{figure}

Our investigation also covered the distribution of the droplet radius. For Weber numbers close to the transition, the radius distribution of the droplets is found bidisperse (Fig. \ref{subfig:distrib_rayon_bidisperse}) with two distinct mean radii. The release of small satellite droplets in addition to the bigger main droplets when a liquid bridge breaks \cite{liCapillaryBreakupLiquid2016,dolganovDynamicsCapillaryCoalescence2021} is a likely explanation for this observation. However, as the Weber number increases (Fig. \ref{subfig:distrib_rayon_monodisperse}), the distribution of the radii of the droplets becomes monodisperse and only one mean radius $r_\mathrm{drop}$ emerges. One possible explanation for this change in distribution is that the conditions conducive to the formation of satellite droplets vanish or simply they become too small to be reliably detected by our code. Henceforth, we will disregard these satellite droplets and focus solely on the largest mean radius $r_\mathrm{drop}$ in the case of bidisperse distributions. It is also noteworthy that the droplets exhibit a radius on the order of $\mathcal{O}(100)$~$\mu$m. This finding aligns with measurements reported in the literature by Oulded Taled Salah \textit{et al.} \cite{ouldedtaledsalahHowTameFree2022} when a Savart sheet breaks.

\begin{figure}[H]
    \centering
    \includegraphics[width=.5\textwidth]{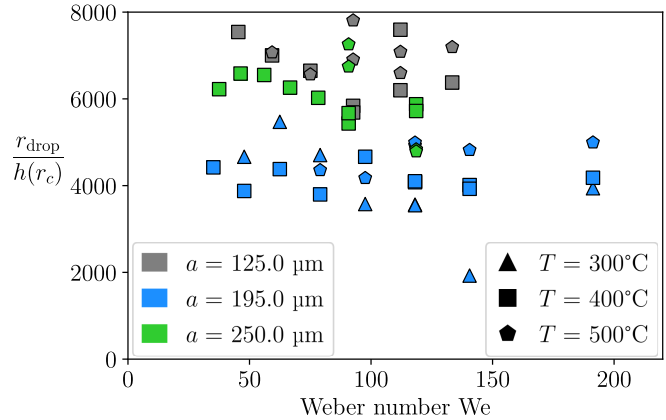} 
    \caption{Ratio between $v_\mathrm{drop}$ and $h(r_c)$ with respect to the Weber number, $\We$.}
    \label{fig:rayon_droplets}
\end{figure}

To get some insights in the physics underlying the size of emerging droplets, we derived the film thickness, $h(r_c)$, at $r_c$ using Eq.~\ref{eq:watson_height_lc_with_ell}, and plotted the ratio $r_\mathrm{drop}/h(r_c)$ as a function of $\We$ in Fig.~\ref{fig:rayon_droplets}.
The results show that the ratio $r_\mathrm{drop}/h(r_c)$ remains nearly constant with $\We$ even though we saw earlier that both $r_\mathrm{drop}$ and $h(r_c)$ vary with $\We$. Furthermore, the plate temperature $T$ seems to have little influence on this ratio, while the effect of the radius $a$ of the impacting jet appears unclear. 

\noindent While the previous scaling of $r_{\text{drop}}$ by $h(r_c)$ appears to be effective in providing insights into the system behavior, it is worth noting another scaling reported in the literature. Clanet \textit{et al.}~\cite{clanetLifeSmoothLiquid2002} nondimensionalized $r_{\text{drop}}$ by $a$ and found the following scaling:
\begin{equation}
    \left(\frac{r_\mathrm{drop}}{a}\right)^3\propto \We^{-1}.
\end{equation}
Our data show a comparable scaling behavior, albeit with a factor of 5 to 10 smaller than that reported by Clanet \textit{et al.} This difference is anticipated because in our system, the heat addition alters and diminishes the effect of surface tension, leading to the production of smaller droplets for the same Weber number, i.e., the same input energy. Moreover, in Clanet \textit{et al.}'s experiment, the liquid-solid contact area is at least five times bigger than in our case. Therefore, the liquid flow experiences less friction in our experiment and retain a considerable speed when the liquid sheet breaks down. However, as the fragmentation must occur when $\Weloc \sim 1$, the liquid sheet's thickness must consequently be smaller in our experiments compared to those of Clanet \textit{et al.}, which leads to the production of smaller droplets.

\begin{figure}[H]
    \centering
    \vskip 0pt
    \includegraphics[width=.5\textwidth]{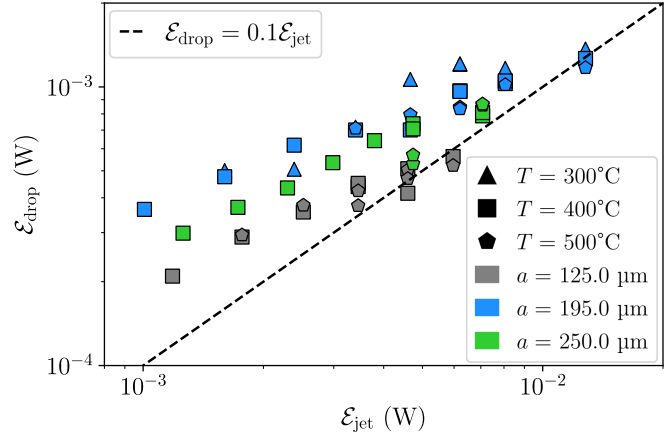} 
    \caption{\justifying Energy taken away by the droplets per unit of time compared to the initial energy of the jet per unit of time. The black dashed line correspond to a 10\% ratio between the two quantities as reported by Clanet \textit{et al.} \cite{clanetLifeSmoothLiquid2002} result.}
    \label{fig:energy}
\end{figure}

To assess the energy taken by the droplets, we followed a similar approach to Clanet \textit{et al.} \cite{clanetLifeSmoothLiquid2002} and considered the energy carried away by the droplets per unit of time as:
\begin{equation}
    \mathcal{E}_\mathrm{drop}=n\left(\frac{2}{3}\pi\rho_{100}r_\mathrm{drop}^3v_\mathrm{drop}^2+4\pi\gamma_{100} r_\mathrm{drop}^2\right),
    \label{eq:e_drop}
\end{equation}
where $n$ is the number of droplets emitted per unit of time. Using the conservation of mass, one finds
\begin{equation}
    n=\frac{3}{4}\frac{a^2}{r_\mathrm{drop}^3}\sqrt{\frac{\gamma \We}{2\rho a}}.
    \label{eq:nb_gouttes}
\end{equation}
Therefore, (\ref{eq:e_drop}) becomes
\begin{equation}
    \mathcal{E}_\mathrm{drop}=\frac{3}{4}\frac{a^2}{r_\mathrm{drop}^3}\sqrt{\frac{\gamma \We}{2\rho a}}\left(\frac{2}{3}\pi\rho_{100}r_\mathrm{drop}^3v_\mathrm{drop}^2+4\pi\gamma_{100} r_\mathrm{drop}^2\right).
    \label{eq:e_drop2}
\end{equation}
As for the initial energy per unit of time of the jet, one can write
\begin{equation}
    \mathcal{E}_\mathrm{jet}=\frac{\pi}{2}\rho a^2\left(\frac{\gamma \We}{2\rho a}\right)^{3/2}.
    \label{eq:e_jet}
\end{equation}
The two energies per unit of time are plotted against each other in Fig. \ref{fig:energy}. In this figure, the black dashed line corresponds to the average experimental ratio reported by Clanet \textit{et al.} \cite{clanetLifeSmoothLiquid2002}. The slope of this line represents the amount of energy transferred from the jet to the droplets or can alternatively be viewed as the energy dissipation that has taken place, as stated by Clanet \textit{et al.} \cite{clanetLifeSmoothLiquid2002}. We prefer to discuss it as an energy transfer where our data provide an indication of how much input energy per unit time $\mathcal{E}_\mathrm{jet}$ is transferred to the energy carried by the droplets $\mathcal{E}_\mathrm{drop}$. When comparing our data to the black dashed line, one can observe that we measure an energy transfer exceeding the 10\% reported by Clanet \textit{et al.}~\cite{clanetLifeSmoothLiquid2002}. This enhancement in energy transfer efficiency is likely attributed to the heat absorbed by the liquid sheet, which modifies the surface tension, thereby facilitating the detachment, and hence, ejection of droplets from the sheet. Of course, this improvement comes at the expense of heat addition to the system, which reduces its overall efficiency. The energy lost in our system can be attributed to heat dissipation, viscous effects, and losses due to evaporation and boiling. Nevertheless, despite these losses, the present system remains of interest if the goal is to dispense droplets on-demand with smaller sizes than what classical systems can achieve, and all this with a fairly simple setup of a liquid impacting a heated surface. One can also wonder what happens for high values of $\mathcal{E}_\mathrm{jet}$, where our data seemingly matches the predictions of Clanet \textit{et al.} \cite{clanetLifeSmoothLiquid2002}. In particular, it may indicate that heat addition to the system has no influence on fast droplets. A more thorough study is however needed to fully address this question.

\section{Conclusion}

We explored the effect of submillimetric liquid jets impacting a heated surface above its Leidenfrost temperature. Two distinct regimes emerged based on the Weber number ($\We$). For $\We \le 30$ the experiments showed the formation of a large centrimetric drops whereas for $\We \geq 40$ the impact yielded to atomization of the jet, leading to distinct physical characteristics such as ejection angle and droplet size. The study mainly focused on characterising the regime showing atomization. While the system displayed weak sensitivity to the temperature of the heated plate and the jet size, simple models successfully explained various scaling characteristics of the system behaviour. Moreover, these models offered explanations for the transition between the regimes, attributed to the interplay between liquid inertia and surface tension. We also conducted detailed measurements of droplet size under various experimental conditions. Our findings indicate an enhanced energy transfer from the jet to droplets compared to systems without surface heating, as supported by existing models and literature data. This system holds promise for practical applications, enabling precise droplet dispensing with smaller sizes than traditional methods, using a relatively straightforward setup of liquid impacting a heated surface.

\section*{Acknowledgements}
The authors acknowledge the support of the French Agence Nationale de la Recherche (ANR), under Grant No. ANR-21-CE30-0014 (project IJET) and thank Loïc Lam for his technical assistance. 

\begin{thebibliography}{37}%
\makeatletter
\providecommand \@ifxundefined [1]{%
 \@ifx{#1\undefined}
}%
\providecommand \@ifnum [1]{%
 \ifnum #1\expandafter \@firstoftwo
 \else \expandafter \@secondoftwo
 \fi
}%
\providecommand \@ifx [1]{%
 \ifx #1\expandafter \@firstoftwo
 \else \expandafter \@secondoftwo
 \fi
}%
\providecommand \natexlab [1]{#1}%
\providecommand \enquote  [1]{``#1''}%
\providecommand \bibnamefont  [1]{#1}%
\providecommand \bibfnamefont [1]{#1}%
\providecommand \citenamefont [1]{#1}%
\providecommand \href@noop [0]{\@secondoftwo}%
\providecommand \href [0]{\begingroup \@sanitize@url \@href}%
\providecommand \@href[1]{\@@startlink{#1}\@@href}%
\providecommand \@@href[1]{\endgroup#1\@@endlink}%
\providecommand \@sanitize@url [0]{\catcode `\\12\catcode `\$12\catcode `\&12\catcode `\#12\catcode `\^12\catcode `\_12\catcode `\%12\relax}%
\providecommand \@@startlink[1]{}%
\providecommand \@@endlink[0]{}%
\providecommand \url  [0]{\begingroup\@sanitize@url \@url }%
\providecommand \@url [1]{\endgroup\@href {#1}{\urlprefix }}%
\providecommand \urlprefix  [0]{URL }%
\providecommand \Eprint [0]{\href }%
\providecommand \doibase [0]{https://doi.org/}%
\providecommand \selectlanguage [0]{\@gobble}%
\providecommand \bibinfo  [0]{\@secondoftwo}%
\providecommand \bibfield  [0]{\@secondoftwo}%
\providecommand \translation [1]{[#1]}%
\providecommand \BibitemOpen [0]{}%
\providecommand \bibitemStop [0]{}%
\providecommand \bibitemNoStop [0]{.\EOS\space}%
\providecommand \EOS [0]{\spacefactor3000\relax}%
\providecommand \BibitemShut  [1]{\csname bibitem#1\endcsname}%
\let\auto@bib@innerbib\@empty
\bibitem [{\citenamefont {Leidenfrost}(1966)}]{leidenfrostFixationWaterDiverse1966}%
  \BibitemOpen
  \bibfield  {author} {\bibinfo {author} {\bibfnamefont {J.~G.}\ \bibnamefont {Leidenfrost}},\ }\bibfield  {title} {\bibinfo {title} {On the fixation of water in diverse fire},\ }\href {https://doi.org/10.1016/0017-9310(66)90111-6} {\bibfield  {journal} {\bibinfo  {journal} {International Journal of Heat and Mass Transfer}\ }\textbf {\bibinfo {volume} {9}},\ \bibinfo {pages} {1153} (\bibinfo {year} {1966})}\BibitemShut {NoStop}%
\bibitem [{\citenamefont {Raudensky}\ and\ \citenamefont {Horsky}(2005)}]{raudenskySecondaryCoolingContinuous2005}%
  \BibitemOpen
  \bibfield  {author} {\bibinfo {author} {\bibfnamefont {M.}~\bibnamefont {Raudensky}}\ and\ \bibinfo {author} {\bibfnamefont {J.}~\bibnamefont {Horsky}},\ }\bibfield  {title} {\bibinfo {title} {Secondary cooling in continuous casting and {{Leidenfrost}} temperature effects},\ }\href {https://doi.org/10.1179/174328105X15913} {\bibfield  {journal} {\bibinfo  {journal} {Ironmaking \& Steelmaking}\ }\textbf {\bibinfo {volume} {32}},\ \bibinfo {pages} {159} (\bibinfo {year} {2005})}\BibitemShut {NoStop}%
\bibitem [{\citenamefont {Ng}\ \emph {et~al.}(2015)\citenamefont {Ng}, \citenamefont {Hung},\ and\ \citenamefont {Tan}}]{ngSuppressionLeidenfrostEffect2015}%
  \BibitemOpen
  \bibfield  {author} {\bibinfo {author} {\bibfnamefont {B.~T.}\ \bibnamefont {Ng}}, \bibinfo {author} {\bibfnamefont {Y.~M.}\ \bibnamefont {Hung}},\ and\ \bibinfo {author} {\bibfnamefont {M.~K.}\ \bibnamefont {Tan}},\ }\bibfield  {title} {\bibinfo {title} {Suppression of the {{Leidenfrost}} effect via low frequency vibrations},\ }\href {https://doi.org/10.1039/C4SM02272F} {\bibfield  {journal} {\bibinfo  {journal} {Soft Matter}\ }\textbf {\bibinfo {volume} {11}},\ \bibinfo {pages} {775} (\bibinfo {year} {2015})}\BibitemShut {NoStop}%
\bibitem [{\citenamefont {Isakov}\ \emph {et~al.}(2017)\citenamefont {Isakov}, \citenamefont {Faber}, \citenamefont {Grell}, \citenamefont {Wyatt-Moon}, \citenamefont {Pliatsikas}, \citenamefont {Kehagias}, \citenamefont {Dimitrakopulos}, \citenamefont {Patsalas}, \citenamefont {Li},\ and\ \citenamefont {Anthopoulos}}]{isakovExploringLeidenfrostEffect2017}%
  \BibitemOpen
  \bibfield  {author} {\bibinfo {author} {\bibfnamefont {I.}~\bibnamefont {Isakov}}, \bibinfo {author} {\bibfnamefont {H.}~\bibnamefont {Faber}}, \bibinfo {author} {\bibfnamefont {M.}~\bibnamefont {Grell}}, \bibinfo {author} {\bibfnamefont {G.}~\bibnamefont {Wyatt-Moon}}, \bibinfo {author} {\bibfnamefont {N.}~\bibnamefont {Pliatsikas}}, \bibinfo {author} {\bibfnamefont {T.}~\bibnamefont {Kehagias}}, \bibinfo {author} {\bibfnamefont {G.~P.}\ \bibnamefont {Dimitrakopulos}}, \bibinfo {author} {\bibfnamefont {P.~P.}\ \bibnamefont {Patsalas}}, \bibinfo {author} {\bibfnamefont {R.}~\bibnamefont {Li}},\ and\ \bibinfo {author} {\bibfnamefont {T.~D.}\ \bibnamefont {Anthopoulos}},\ }\bibfield  {title} {\bibinfo {title} {Exploring the {{Leidenfrost Effect}} for the {{Deposition}} of {{High}}-{{Quality In}}{\textsubscript{2}}{{O}}{\textsubscript{3}} {{Layers}} via {{Spray Pyrolysis}} at {{Low Temperatures}} and {{Their Application}} in {{High Electron Mobility Transistors}}},\ }\href
  {https://doi.org/10.1002/adfm.201606407} {\bibfield  {journal} {\bibinfo  {journal} {Advanced Functional Materials}\ }\textbf {\bibinfo {volume} {27}},\ \bibinfo {pages} {1606407} (\bibinfo {year} {2017})}\BibitemShut {NoStop}%
\bibitem [{\citenamefont {Bernardin}\ and\ \citenamefont {Mudawar}(1999)}]{bernardinLeidenfrostPointExperimental1999}%
  \BibitemOpen
  \bibfield  {author} {\bibinfo {author} {\bibfnamefont {J.~D.}\ \bibnamefont {Bernardin}}\ and\ \bibinfo {author} {\bibfnamefont {I.}~\bibnamefont {Mudawar}},\ }\bibfield  {title} {\bibinfo {title} {The {{Leidenfrost Point}}: {{Experimental Study}} and {{Assessment}} of {{Existing Models}}},\ }\href {https://doi.org/10.1115/1.2826080} {\bibfield  {journal} {\bibinfo  {journal} {Journal of Heat Transfer}\ }\textbf {\bibinfo {volume} {121}},\ \bibinfo {pages} {894} (\bibinfo {year} {1999})}\BibitemShut {NoStop}%
\bibitem [{\citenamefont {Qu{\'e}r{\'e}}(2013)}]{quereLeidenfrostDynamics2013}%
  \BibitemOpen
  \bibfield  {author} {\bibinfo {author} {\bibfnamefont {D.}~\bibnamefont {Qu{\'e}r{\'e}}},\ }\bibfield  {title} {\bibinfo {title} {Leidenfrost {{Dynamics}}},\ }\href {https://doi.org/10.1146/annurev-fluid-011212-140709} {\bibfield  {journal} {\bibinfo  {journal} {Annual Review of Fluid Mechanics}\ }\textbf {\bibinfo {volume} {45}},\ \bibinfo {pages} {197} (\bibinfo {year} {2013})}\BibitemShut {NoStop}%
\bibitem [{\citenamefont {Liang}\ and\ \citenamefont {Mudawar}(2017)}]{liangReviewDropImpact2017}%
  \BibitemOpen
  \bibfield  {author} {\bibinfo {author} {\bibfnamefont {G.}~\bibnamefont {Liang}}\ and\ \bibinfo {author} {\bibfnamefont {I.}~\bibnamefont {Mudawar}},\ }\bibfield  {title} {\bibinfo {title} {Review of drop impact on heated walls},\ }\href {https://doi.org/10.1016/j.ijheatmasstransfer.2016.10.031} {\bibfield  {journal} {\bibinfo  {journal} {International Journal of Heat and Mass Transfer}\ }\textbf {\bibinfo {volume} {106}},\ \bibinfo {pages} {103} (\bibinfo {year} {2017})}\BibitemShut {NoStop}%
\bibitem [{\citenamefont {Cai}\ and\ \citenamefont {Mudawar}(2023)}]{caiReviewDynamicLeidenfrost2023}%
  \BibitemOpen
  \bibfield  {author} {\bibinfo {author} {\bibfnamefont {C.}~\bibnamefont {Cai}}\ and\ \bibinfo {author} {\bibfnamefont {I.}~\bibnamefont {Mudawar}},\ }\bibfield  {title} {\bibinfo {title} {Review of the dynamic {{Leidenfrost}} point temperature for droplet impact on a heated solid surface},\ }\href {https://doi.org/10.1016/j.ijheatmasstransfer.2023.124639} {\bibfield  {journal} {\bibinfo  {journal} {International Journal of Heat and Mass Transfer}\ }\textbf {\bibinfo {volume} {217}},\ \bibinfo {pages} {124639} (\bibinfo {year} {2023})}\BibitemShut {NoStop}%
\bibitem [{\citenamefont {Lastakowski}\ \emph {et~al.}(2014)\citenamefont {Lastakowski}, \citenamefont {Boyer}, \citenamefont {Biance}, \citenamefont {Pirat},\ and\ \citenamefont {Ybert}}]{lastakowskiBridgingLocalGlobal2014}%
  \BibitemOpen
  \bibfield  {author} {\bibinfo {author} {\bibfnamefont {H.}~\bibnamefont {Lastakowski}}, \bibinfo {author} {\bibfnamefont {F.}~\bibnamefont {Boyer}}, \bibinfo {author} {\bibfnamefont {A.-L.}\ \bibnamefont {Biance}}, \bibinfo {author} {\bibfnamefont {C.}~\bibnamefont {Pirat}},\ and\ \bibinfo {author} {\bibfnamefont {C.}~\bibnamefont {Ybert}},\ }\bibfield  {title} {\bibinfo {title} {Bridging local to global dynamics of drop impact onto solid substrates},\ }\href {https://doi.org/10.1017/jfm.2014.108} {\bibfield  {journal} {\bibinfo  {journal} {Journal of Fluid Mechanics}\ }\textbf {\bibinfo {volume} {747}},\ \bibinfo {pages} {103} (\bibinfo {year} {2014})}\BibitemShut {NoStop}%
\bibitem [{\citenamefont {Riboux}\ and\ \citenamefont {Gordillo}(2016)}]{ribouxMaximumDropRadius2016}%
  \BibitemOpen
  \bibfield  {author} {\bibinfo {author} {\bibfnamefont {G.}~\bibnamefont {Riboux}}\ and\ \bibinfo {author} {\bibfnamefont {J.~M.}\ \bibnamefont {Gordillo}},\ }\bibfield  {title} {\bibinfo {title} {Maximum drop radius and critical {{Weber}} number for splashing in the dynamical {{Leidenfrost}} regime},\ }\href {https://doi.org/10.1017/jfm.2016.496} {\bibfield  {journal} {\bibinfo  {journal} {Journal of Fluid Mechanics}\ }\textbf {\bibinfo {volume} {803}},\ \bibinfo {pages} {516} (\bibinfo {year} {2016})}\BibitemShut {NoStop}%
\bibitem [{\citenamefont {Biance}\ \emph {et~al.}(2006)\citenamefont {Biance}, \citenamefont {Chevy}, \citenamefont {Clanet}, \citenamefont {Lagubeau},\ and\ \citenamefont {Qu{\'e}r{\'e}}}]{bianceElasticityInertialLiquid2006}%
  \BibitemOpen
  \bibfield  {author} {\bibinfo {author} {\bibfnamefont {A.-L.}\ \bibnamefont {Biance}}, \bibinfo {author} {\bibfnamefont {F.}~\bibnamefont {Chevy}}, \bibinfo {author} {\bibfnamefont {C.}~\bibnamefont {Clanet}}, \bibinfo {author} {\bibfnamefont {G.}~\bibnamefont {Lagubeau}},\ and\ \bibinfo {author} {\bibfnamefont {D.}~\bibnamefont {Qu{\'e}r{\'e}}},\ }\bibfield  {title} {\bibinfo {title} {On the elasticity of an inertial liquid shock},\ }\href {https://doi.org/10.1017/S0022112006009189} {\bibfield  {journal} {\bibinfo  {journal} {Journal of Fluid Mechanics}\ }\textbf {\bibinfo {volume} {554}},\ \bibinfo {pages} {47} (\bibinfo {year} {2006})}\BibitemShut {NoStop}%
\bibitem [{\citenamefont {Biance}\ \emph {et~al.}(2011)\citenamefont {Biance}, \citenamefont {Pirat},\ and\ \citenamefont {Ybert}}]{bianceDropFragmentationDue2011}%
  \BibitemOpen
  \bibfield  {author} {\bibinfo {author} {\bibfnamefont {A.-L.}\ \bibnamefont {Biance}}, \bibinfo {author} {\bibfnamefont {C.}~\bibnamefont {Pirat}},\ and\ \bibinfo {author} {\bibfnamefont {C.}~\bibnamefont {Ybert}},\ }\bibfield  {title} {\bibinfo {title} {Drop fragmentation due to hole formation during {{Leidenfrost}} impact},\ }\href {https://doi.org/10.1063/1.3553277} {\bibfield  {journal} {\bibinfo  {journal} {Physics of Fluids}\ }\textbf {\bibinfo {volume} {23}},\ \bibinfo {pages} {022104} (\bibinfo {year} {2011})}\BibitemShut {NoStop}%
\bibitem [{\citenamefont {{van Limbeek}}\ \emph {et~al.}(2017)\citenamefont {{van Limbeek}}, \citenamefont {Hoefnagels}, \citenamefont {Sun},\ and\ \citenamefont {Lohse}}]{vanlimbeekOriginSprayFormation2017}%
  \BibitemOpen
  \bibfield  {author} {\bibinfo {author} {\bibfnamefont {M.~A.~J.}\ \bibnamefont {{van Limbeek}}}, \bibinfo {author} {\bibfnamefont {P.~B.~J.}\ \bibnamefont {Hoefnagels}}, \bibinfo {author} {\bibfnamefont {C.}~\bibnamefont {Sun}},\ and\ \bibinfo {author} {\bibfnamefont {D.}~\bibnamefont {Lohse}},\ }\bibfield  {title} {\bibinfo {title} {Origin of spray formation during impact on heated surfaces},\ }\href {https://doi.org/10.1039/C7SM00956A} {\bibfield  {journal} {\bibinfo  {journal} {Soft Matter}\ }\textbf {\bibinfo {volume} {13}},\ \bibinfo {pages} {7514} (\bibinfo {year} {2017})}\BibitemShut {NoStop}%
\bibitem [{\citenamefont {Tran}\ \emph {et~al.}(2012)\citenamefont {Tran}, \citenamefont {Staat}, \citenamefont {Prosperetti}, \citenamefont {Sun},\ and\ \citenamefont {Lohse}}]{tranDropImpactSuperheated2012}%
  \BibitemOpen
  \bibfield  {author} {\bibinfo {author} {\bibfnamefont {T.}~\bibnamefont {Tran}}, \bibinfo {author} {\bibfnamefont {H.~J.~J.}\ \bibnamefont {Staat}}, \bibinfo {author} {\bibfnamefont {A.}~\bibnamefont {Prosperetti}}, \bibinfo {author} {\bibfnamefont {C.}~\bibnamefont {Sun}},\ and\ \bibinfo {author} {\bibfnamefont {D.}~\bibnamefont {Lohse}},\ }\bibfield  {title} {\bibinfo {title} {Drop {{Impact}} on {{Superheated Surfaces}}},\ }\href {https://doi.org/10.1103/PhysRevLett.108.036101} {\bibfield  {journal} {\bibinfo  {journal} {Physical Review Letters}\ }\textbf {\bibinfo {volume} {108}},\ \bibinfo {pages} {036101} (\bibinfo {year} {2012})}\BibitemShut {NoStop}%
\bibitem [{\citenamefont {Khavari}\ \emph {et~al.}(2015)\citenamefont {Khavari}, \citenamefont {Sun}, \citenamefont {Lohse},\ and\ \citenamefont {Tran}}]{khavariFingeringPatternsDroplet2015}%
  \BibitemOpen
  \bibfield  {author} {\bibinfo {author} {\bibfnamefont {M.}~\bibnamefont {Khavari}}, \bibinfo {author} {\bibfnamefont {C.}~\bibnamefont {Sun}}, \bibinfo {author} {\bibfnamefont {D.}~\bibnamefont {Lohse}},\ and\ \bibinfo {author} {\bibfnamefont {T.}~\bibnamefont {Tran}},\ }\bibfield  {title} {\bibinfo {title} {Fingering patterns during droplet impact on heated surfaces},\ }\href {https://doi.org/10.1039/C4SM02878C} {\bibfield  {journal} {\bibinfo  {journal} {Soft Matter}\ }\textbf {\bibinfo {volume} {11}},\ \bibinfo {pages} {3298} (\bibinfo {year} {2015})}\BibitemShut {NoStop}%
\bibitem [{\citenamefont {Agrawal}(2019)}]{agrawalSurfaceQuenchingJet2019}%
  \BibitemOpen
  \bibfield  {author} {\bibinfo {author} {\bibfnamefont {C.}~\bibnamefont {Agrawal}},\ }\bibfield  {title} {\bibinfo {title} {Surface {{Quenching}} by {{Jet Impingement}} - {{A Review}}},\ }\href {https://doi.org/10.1002/srin.201800285} {\bibfield  {journal} {\bibinfo  {journal} {steel research international}\ }\textbf {\bibinfo {volume} {90}},\ \bibinfo {pages} {1800285} (\bibinfo {year} {2019})}\BibitemShut {NoStop}%
\bibitem [{\citenamefont {Vader}(1988)}]{vaderConvectiveBoilingHeat1988}%
  \BibitemOpen
  \bibfield  {author} {\bibinfo {author} {\bibfnamefont {D.~T.}\ \bibnamefont {Vader}},\ }\emph {\bibinfo {title} {Convective Boiling Heat Transfer from a Heated Surface to an Impinging, Planar Jet of Water}},\ \href@noop {} {Ph.D. thesis},\ \bibinfo  {school} {Purdue University} (\bibinfo {year} {1988})\BibitemShut {NoStop}%
\bibitem [{\citenamefont {Wolf}(1993)}]{wolfTurbulentDevelopmentFree1993}%
  \BibitemOpen
  \bibfield  {author} {\bibinfo {author} {\bibfnamefont {D.~H.}\ \bibnamefont {Wolf}},\ }\emph {\bibinfo {title} {Turbulent Development in a Free Surface Jet and Impingement Boiling Heat Transfer}},\ \href@noop {} {Ph.D. thesis},\ \bibinfo  {school} {Purdue university} (\bibinfo {year} {1993})\BibitemShut {NoStop}%
\bibitem [{\citenamefont {Robidou}\ \emph {et~al.}(2002)\citenamefont {Robidou}, \citenamefont {Auracher}, \citenamefont {Gardin},\ and\ \citenamefont {Lebouch{\'e}}}]{robidouControlledCoolingHot2002}%
  \BibitemOpen
  \bibfield  {author} {\bibinfo {author} {\bibfnamefont {H.}~\bibnamefont {Robidou}}, \bibinfo {author} {\bibfnamefont {H.}~\bibnamefont {Auracher}}, \bibinfo {author} {\bibfnamefont {P.}~\bibnamefont {Gardin}},\ and\ \bibinfo {author} {\bibfnamefont {M.}~\bibnamefont {Lebouch{\'e}}},\ }\bibfield  {title} {\bibinfo {title} {Controlled cooling of a hot plate with a water jet},\ }\href {https://doi.org/10.1016/S0894-1777(02)00118-8} {\bibfield  {journal} {\bibinfo  {journal} {Experimental Thermal and Fluid Science}\ }\textbf {\bibinfo {volume} {26}},\ \bibinfo {pages} {123} (\bibinfo {year} {2002})}\BibitemShut {NoStop}%
\bibitem [{\citenamefont {Robidou}\ \emph {et~al.}(2003)\citenamefont {Robidou}, \citenamefont {Auracher}, \citenamefont {Gardin}, \citenamefont {Lebouch{\'e}},\ and\ \citenamefont {Bogdanic}}]{robidouLocalHeatTransfer2003}%
  \BibitemOpen
  \bibfield  {author} {\bibinfo {author} {\bibfnamefont {H.}~\bibnamefont {Robidou}}, \bibinfo {author} {\bibfnamefont {H.}~\bibnamefont {Auracher}}, \bibinfo {author} {\bibfnamefont {P.}~\bibnamefont {Gardin}}, \bibinfo {author} {\bibfnamefont {M.}~\bibnamefont {Lebouch{\'e}}},\ and\ \bibinfo {author} {\bibfnamefont {L.}~\bibnamefont {Bogdanic}},\ }\bibfield  {title} {\bibinfo {title} {Local heat transfer from a hot plate to a water jet},\ }\href {https://doi.org/10.1007/s00231-002-0335-6} {\bibfield  {journal} {\bibinfo  {journal} {Heat and Mass Transfer}\ }\textbf {\bibinfo {volume} {39}},\ \bibinfo {pages} {861} (\bibinfo {year} {2003})}\BibitemShut {NoStop}%
\bibitem [{\citenamefont {Bogdanic}\ \emph {et~al.}(2009)\citenamefont {Bogdanic}, \citenamefont {Auracher},\ and\ \citenamefont {Ziegler}}]{bogdanicTwophaseStructureHot2009}%
  \BibitemOpen
  \bibfield  {author} {\bibinfo {author} {\bibfnamefont {L.}~\bibnamefont {Bogdanic}}, \bibinfo {author} {\bibfnamefont {H.}~\bibnamefont {Auracher}},\ and\ \bibinfo {author} {\bibfnamefont {F.}~\bibnamefont {Ziegler}},\ }\bibfield  {title} {\bibinfo {title} {Two-phase structure above hot surfaces in jet impingement boiling},\ }\href {https://doi.org/10.1007/s00231-007-0272-5} {\bibfield  {journal} {\bibinfo  {journal} {Heat and Mass Transfer}\ }\textbf {\bibinfo {volume} {45}},\ \bibinfo {pages} {1019} (\bibinfo {year} {2009})}\BibitemShut {NoStop}%
\bibitem [{\citenamefont {K{\"o}berle}\ and\ \citenamefont {Auracher}(1994)}]{koberleTemperatureControlledMeasurements1994}%
  \BibitemOpen
  \bibfield  {author} {\bibinfo {author} {\bibfnamefont {K.}~\bibnamefont {K{\"o}berle}}\ and\ \bibinfo {author} {\bibfnamefont {H.}~\bibnamefont {Auracher}},\ }\bibfield  {title} {\bibinfo {title} {Temperature controlled measurements of the critical heat flux on microelectronic heat sources in natural convection and jet impingement cooling},\ }in\ \href {https://doi.org/10.1007/978-94-011-1082-2_21} {\emph {\bibinfo {booktitle} {Thermal {{Management}} of {{Electronic Systems}}}}},\ \bibinfo {editor} {edited by\ \bibinfo {editor} {\bibfnamefont {C.~J.}\ \bibnamefont {Hoogendoorn}}, \bibinfo {editor} {\bibfnamefont {R.~A. W.~M.}\ \bibnamefont {Henkes}},\ and\ \bibinfo {editor} {\bibfnamefont {C.~J.~M.}\ \bibnamefont {Lasance}}}\ (\bibinfo  {publisher} {{Springer Netherlands}},\ \bibinfo {address} {{Dordrecht}},\ \bibinfo {year} {1994})\ pp.\ \bibinfo {pages} {233--242}\BibitemShut {NoStop}%
\bibitem [{\citenamefont {Baonga}\ \emph {et~al.}(2006)\citenamefont {Baonga}, \citenamefont {{Louahlia-Gualous}},\ and\ \citenamefont {Imbert}}]{baongaExperimentalStudyHydrodynamic2006}%
  \BibitemOpen
  \bibfield  {author} {\bibinfo {author} {\bibfnamefont {J.~B.}\ \bibnamefont {Baonga}}, \bibinfo {author} {\bibfnamefont {H.}~\bibnamefont {{Louahlia-Gualous}}},\ and\ \bibinfo {author} {\bibfnamefont {M.}~\bibnamefont {Imbert}},\ }\bibfield  {title} {\bibinfo {title} {Experimental study of the hydrodynamic and heat transfer of free liquid jet impinging a at circular heated disk},\ }\href@noop {} {\bibfield  {journal} {\bibinfo  {journal} {Applied Thermal Engineering}\ }\textbf {\bibinfo {volume} {26}},\ \bibinfo {pages} {1125} (\bibinfo {year} {2006})}\BibitemShut {NoStop}%
\bibitem [{\citenamefont {Leocadio}\ \emph {et~al.}(2009)\citenamefont {Leocadio}, \citenamefont {Passos},\ and\ \citenamefont {Silva}}]{leocadioHeatTransferBehavior2009}%
  \BibitemOpen
  \bibfield  {author} {\bibinfo {author} {\bibfnamefont {H.}~\bibnamefont {Leocadio}}, \bibinfo {author} {\bibfnamefont {J.~C.}\ \bibnamefont {Passos}},\ and\ \bibinfo {author} {\bibfnamefont {A.~F. C.~D.}\ \bibnamefont {Silva}},\ }\bibfield  {title} {\bibinfo {title} {Heat transfer behavior of a high temperature steel plate cooled by a subcooled impinging circular water jet},\ }\href@noop {} {\bibfield  {journal} {\bibinfo  {journal} {7th International Conference on Boiling Heat Transfer}\ ,\ \bibinfo {pages} {3}} (\bibinfo {year} {2009})}\BibitemShut {NoStop}%
\bibitem [{\citenamefont {Karwa}\ \emph {et~al.}(2011)\citenamefont {Karwa}, \citenamefont {{Gambaryan-Roisman}}, \citenamefont {Stephan},\ and\ \citenamefont {Tropea}}]{karwaHydrodynamicModelSubcooled2011}%
  \BibitemOpen
  \bibfield  {author} {\bibinfo {author} {\bibfnamefont {N.}~\bibnamefont {Karwa}}, \bibinfo {author} {\bibfnamefont {T.}~\bibnamefont {{Gambaryan-Roisman}}}, \bibinfo {author} {\bibfnamefont {P.}~\bibnamefont {Stephan}},\ and\ \bibinfo {author} {\bibfnamefont {C.}~\bibnamefont {Tropea}},\ }\bibfield  {title} {\bibinfo {title} {A hydrodynamic model for subcooled liquid jet impingement at the {{Leidenfrost}} condition},\ }\href {https://doi.org/10.1016/j.ijthermalsci.2011.01.021} {\bibfield  {journal} {\bibinfo  {journal} {International Journal of Thermal Sciences}\ }\textbf {\bibinfo {volume} {50}},\ \bibinfo {pages} {993} (\bibinfo {year} {2011})}\BibitemShut {NoStop}%
\bibitem [{\citenamefont {Ma}\ \emph {et~al.}(1997)\citenamefont {Ma}, \citenamefont {Zheng}, \citenamefont {Sun}, \citenamefont {Wu}, \citenamefont {Gomi},\ and\ \citenamefont {Webb}}]{maLocalCharacteristicsImpingement1997a}%
  \BibitemOpen
  \bibfield  {author} {\bibinfo {author} {\bibfnamefont {C.}~\bibnamefont {Ma}}, \bibinfo {author} {\bibfnamefont {Q.}~\bibnamefont {Zheng}}, \bibinfo {author} {\bibfnamefont {H.}~\bibnamefont {Sun}}, \bibinfo {author} {\bibfnamefont {K.}~\bibnamefont {Wu}}, \bibinfo {author} {\bibfnamefont {T.}~\bibnamefont {Gomi}},\ and\ \bibinfo {author} {\bibfnamefont {B.}~\bibnamefont {Webb}},\ }\bibfield  {title} {\bibinfo {title} {Local characteristics of impingement heat transfer with oblique round free-surface jets of large {{Prandtl}} number liquid},\ }\href {https://doi.org/10.1016/S0017-9310(96)00310-9} {\bibfield  {journal} {\bibinfo  {journal} {International Journal of Heat and Mass Transfer}\ }\textbf {\bibinfo {volume} {40}},\ \bibinfo {pages} {2249} (\bibinfo {year} {1997})}\BibitemShut {NoStop}%
\bibitem [{Sup()}]{SuppMat}%
  \BibitemOpen
  \href@noop {} {\bibinfo {title} {See {Supplementary Materials} at for movies {M1} and {M2}.}}\BibitemShut {Stop}%
\bibitem [{\citenamefont {Burzynski}\ \emph {et~al.}(2020)\citenamefont {Burzynski}, \citenamefont {Roisman},\ and\ \citenamefont {Bansmer}}]{burzynskiSplashingHighspeedDrops2020}%
  \BibitemOpen
  \bibfield  {author} {\bibinfo {author} {\bibfnamefont {D.~A.}\ \bibnamefont {Burzynski}}, \bibinfo {author} {\bibfnamefont {I.~V.}\ \bibnamefont {Roisman}},\ and\ \bibinfo {author} {\bibfnamefont {S.~E.}\ \bibnamefont {Bansmer}},\ }\bibfield  {title} {\bibinfo {title} {On the splashing of high-speed drops impacting a dry surface},\ }\href {https://doi.org/10.1017/jfm.2020.168} {\bibfield  {journal} {\bibinfo  {journal} {Journal of Fluid Mechanics}\ }\textbf {\bibinfo {volume} {892}},\ \bibinfo {pages} {A2} (\bibinfo {year} {2020})}\BibitemShut {NoStop}%
\bibitem [{\citenamefont {Lin}\ and\ \citenamefont {Jiang}(2003)}]{linAbsoluteConvectiveInstability2003}%
  \BibitemOpen
  \bibfield  {author} {\bibinfo {author} {\bibfnamefont {S.~P.}\ \bibnamefont {Lin}}\ and\ \bibinfo {author} {\bibfnamefont {W.~Y.}\ \bibnamefont {Jiang}},\ }\bibfield  {title} {\bibinfo {title} {Absolute and convective instability of a radially expanding liquid sheet},\ }\href {https://doi.org/10.1063/1.1570422} {\bibfield  {journal} {\bibinfo  {journal} {Physics of Fluids}\ }\textbf {\bibinfo {volume} {15}},\ \bibinfo {pages} {1745} (\bibinfo {year} {2003})}\BibitemShut {NoStop}%
\bibitem [{\citenamefont {Maynes}\ \emph {et~al.}(2011)\citenamefont {Maynes}, \citenamefont {Johnson},\ and\ \citenamefont {Webb}}]{maynesFreesurfaceLiquidJet2011}%
  \BibitemOpen
  \bibfield  {author} {\bibinfo {author} {\bibfnamefont {D.}~\bibnamefont {Maynes}}, \bibinfo {author} {\bibfnamefont {M.}~\bibnamefont {Johnson}},\ and\ \bibinfo {author} {\bibfnamefont {B.~W.}\ \bibnamefont {Webb}},\ }\bibfield  {title} {\bibinfo {title} {Free-surface liquid jet impingement on rib patterned superhydrophobic surfaces},\ }\href@noop {} {\bibfield  {journal} {\bibinfo  {journal} {Physics of Fluids}\ }\textbf {\bibinfo {volume} {23}} (\bibinfo {year} {2011})}\BibitemShut {NoStop}%
\bibitem [{\citenamefont {Watson}(1964)}]{watsonRadialSpreadLiquid1964}%
  \BibitemOpen
  \bibfield  {author} {\bibinfo {author} {\bibfnamefont {E.~J.}\ \bibnamefont {Watson}},\ }\bibfield  {title} {\bibinfo {title} {The radial spread of a liquid jet over a horizontal plane},\ }\href@noop {} {\bibfield  {journal} {\bibinfo  {journal} {Journal of Fluid Mechanics}\ }\textbf {\bibinfo {volume} {20}},\ \bibinfo {pages} {481} (\bibinfo {year} {1964})}\BibitemShut {NoStop}%
\bibitem [{\citenamefont {Celestini}\ \emph {et~al.}(2012)\citenamefont {Celestini}, \citenamefont {Frisch},\ and\ \citenamefont {Pomeau}}]{celestiniTakeSmallLeidenfrost2012}%
  \BibitemOpen
  \bibfield  {author} {\bibinfo {author} {\bibfnamefont {F.}~\bibnamefont {Celestini}}, \bibinfo {author} {\bibfnamefont {T.}~\bibnamefont {Frisch}},\ and\ \bibinfo {author} {\bibfnamefont {Y.}~\bibnamefont {Pomeau}},\ }\bibfield  {title} {\bibinfo {title} {Take {{Off}} of {{Small Leidenfrost Droplets}}},\ }\href {https://doi.org/10.1103/PhysRevLett.109.034501} {\bibfield  {journal} {\bibinfo  {journal} {Physical Review Letters}\ }\textbf {\bibinfo {volume} {109}},\ \bibinfo {pages} {034501} (\bibinfo {year} {2012})}\BibitemShut {NoStop}%
\bibitem [{\citenamefont {Clanet}\ and\ \citenamefont {Villermaux}(2002)}]{clanetLifeSmoothLiquid2002}%
  \BibitemOpen
  \bibfield  {author} {\bibinfo {author} {\bibfnamefont {C.}~\bibnamefont {Clanet}}\ and\ \bibinfo {author} {\bibfnamefont {E.}~\bibnamefont {Villermaux}},\ }\bibfield  {title} {\bibinfo {title} {Life of a smooth liquid sheet},\ }\href {https://doi.org/10.1017/S0022112002008339} {\bibfield  {journal} {\bibinfo  {journal} {Journal of Fluid Mechanics}\ }\textbf {\bibinfo {volume} {462}},\ \bibinfo {pages} {307} (\bibinfo {year} {2002})}\BibitemShut {NoStop}%
\bibitem [{\citenamefont {Crocker}\ and\ \citenamefont {Grier}(1996)}]{crockerMethodsDigitalVideo1996}%
  \BibitemOpen
  \bibfield  {author} {\bibinfo {author} {\bibfnamefont {J.~C.}\ \bibnamefont {Crocker}}\ and\ \bibinfo {author} {\bibfnamefont {D.~G.}\ \bibnamefont {Grier}},\ }\bibfield  {title} {\bibinfo {title} {Methods of {{Digital Video Microscopy}} for {{Colloidal Studies}}},\ }\href {https://doi.org/10.1006/jcis.1996.0217} {\bibfield  {journal} {\bibinfo  {journal} {Journal of Colloid and Interface Science}\ }\textbf {\bibinfo {volume} {179}},\ \bibinfo {pages} {298} (\bibinfo {year} {1996})}\BibitemShut {NoStop}%
\bibitem [{\citenamefont {Li}\ and\ \citenamefont {Sprittles}(2016)}]{liCapillaryBreakupLiquid2016}%
  \BibitemOpen
  \bibfield  {author} {\bibinfo {author} {\bibfnamefont {Y.}~\bibnamefont {Li}}\ and\ \bibinfo {author} {\bibfnamefont {J.~E.}\ \bibnamefont {Sprittles}},\ }\bibfield  {title} {\bibinfo {title} {Capillary breakup of a liquid bridge: Identifying regimes and transitions},\ }\href {https://doi.org/10.1017/jfm.2016.276} {\bibfield  {journal} {\bibinfo  {journal} {Journal of Fluid Mechanics}\ }\textbf {\bibinfo {volume} {797}},\ \bibinfo {pages} {29} (\bibinfo {year} {2016})}\BibitemShut {NoStop}%
\bibitem [{\citenamefont {Dolganov}\ \emph {et~al.}(2021)\citenamefont {Dolganov}, \citenamefont {Zverev}, \citenamefont {Baklanova},\ and\ \citenamefont {Dolganov}}]{dolganovDynamicsCapillaryCoalescence2021}%
  \BibitemOpen
  \bibfield  {author} {\bibinfo {author} {\bibfnamefont {P.~V.}\ \bibnamefont {Dolganov}}, \bibinfo {author} {\bibfnamefont {A.~S.}\ \bibnamefont {Zverev}}, \bibinfo {author} {\bibfnamefont {K.~D.}\ \bibnamefont {Baklanova}},\ and\ \bibinfo {author} {\bibfnamefont {V.~K.}\ \bibnamefont {Dolganov}},\ }\bibfield  {title} {\bibinfo {title} {Dynamics of capillary coalescence and breakup: {{Quasi-two-dimensional}} nematic and isotropic droplets},\ }\href {https://doi.org/10.1103/PhysRevE.104.014702} {\bibfield  {journal} {\bibinfo  {journal} {Physical Review E}\ }\textbf {\bibinfo {volume} {104}},\ \bibinfo {pages} {014702} (\bibinfo {year} {2021})}\BibitemShut {NoStop}%
\bibitem [{\citenamefont {Oulded Taled~Salah}\ \emph {et~al.}(2022)\citenamefont {Oulded Taled~Salah}, \citenamefont {Chouk}, \citenamefont {Duchesne}, \citenamefont {De~Cock}, \citenamefont {Abrougui}, \citenamefont {Lebeau},\ and\ \citenamefont {Dorbolo}}]{ouldedtaledsalahHowTameFree2022}%
  \BibitemOpen
  \bibfield  {author} {\bibinfo {author} {\bibfnamefont {S.}~\bibnamefont {Oulded Taled~Salah}}, \bibinfo {author} {\bibfnamefont {G.}~\bibnamefont {Chouk}}, \bibinfo {author} {\bibfnamefont {A.}~\bibnamefont {Duchesne}}, \bibinfo {author} {\bibfnamefont {N.}~\bibnamefont {De~Cock}}, \bibinfo {author} {\bibfnamefont {K.}~\bibnamefont {Abrougui}}, \bibinfo {author} {\bibfnamefont {F.}~\bibnamefont {Lebeau}},\ and\ \bibinfo {author} {\bibfnamefont {S.}~\bibnamefont {Dorbolo}},\ }\bibfield  {title} {\bibinfo {title} {How to tame a free non-laminar {{Savart}} sheet into individual jets?},\ }\href {https://doi.org/10.1016/j.ijmultiphaseflow.2022.104032} {\bibfield  {journal} {\bibinfo  {journal} {International Journal of Multiphase Flow}\ }\textbf {\bibinfo {volume} {152}},\ \bibinfo {pages} {104032} (\bibinfo {year} {2022})}\BibitemShut {NoStop}%
\end{thebibliography}

\providecommand{\noopsort}[1]{}\providecommand{\singleletter}[1]{#1}%

\end{document}